\documentclass[twocolumn,aps,nofootinbib,epsfig]{revtex4}
\usepackage{graphicx}
\usepackage{dcolumn}
\usepackage{bm}
\usepackage{amssymb}
\usepackage{siunitx}
\usepackage{amsmath}
\usepackage{graphicx}
\usepackage{multirow} 
\usepackage{epsfig}
\usepackage{caption}
\usepackage[utf8]{inputenc} 
\usepackage{ucs} 
\usepackage{amsmath} 
\usepackage{array}
\usepackage{float}
\usepackage{color}
\usepackage[usenames,dvipsnames]{xcolor}
\usepackage{epstopdf}
\usepackage{natbib}
\usepackage{graphicx}
\usepackage{lineno}
\usepackage{slashed}
\usepackage{color}
\usepackage{caption}
\usepackage{hyperref}
\definecolor{red}{rgb}{1.0,0.0,0.0}
\definecolor{blue}{rgb}{0.0,0.0,1}
\newcommand{\T}{\tilde}
\newcommand{\n}{\mathbf}
\newcommand{\bea}{\begin{eqnarray}}
\newcommand{\eea}{\end{eqnarray}}
\newcommand{\nn}{\nonumber}
\newcommand{\etal}{\textit{et al.}}

\newcommand{\nmn}{{\eta^\mu}_\nu}

\usepackage{soul}
\begin{document}
%
%
\title{Optical intersubband properties of a core-shell semiconductor-topological insulator quantum dot described by $\theta$-Electrodynamics}

%
%
\author{Jorge David Casta\~no-Yepes$^1$\footnote{Corresponding author.\\
E-mail address: jorgecastanoy@gmail.com (J.D. Castaño-Yepes).}, O. J. Franca$^1$, C. F. Ramirez-Gutierrez$^{2}$ and J. C. del Valle$^1$}
\address{
  $^1$Instituto de Ciencias
  Nucleares, Universidad Nacional Aut\'onoma de M\'exico, Apartado
  Postal 70-543, Ciudad de M\'exico 04510, M\'exico.\\
  $^2$Universidad Polit\'ecnica de Quer\'etaro, C.P. 76240,  El Marqu\'es, Qro., M\'exico}
%
\begin{abstract}
The optical properties of a spherical topological insulator embedded concentrically in a single-electron system consisting of a core-shell GaAs quantum dot are analysed, when the system is under a uniform external magnetic field. The topological insulator's magnetoelectric response is computed in the effective framework of $\theta$-electrodynamics, which allows analytical calculations for the induced electric and magnetic potentials by Green's Function method. The GaAs  Hamiltonian is constructed in the effective-mass approximation, and its corresponding Schrödinger equation is numerically solved through the Lagrange-mesh method. We compute the total absorption coefficients and refractive index changes given by the non-linear iterative density matrix formalism up to third order. Our results show that the presence of the magnetoelectric material causes new dipolar transitions otherwise not allowed. Also, an enhancement of the photon absorption is found when the incident light polarization is oriented parallel to the external magnetic field, in comparison with perpendicular polarization. Moreover, we report an appreciable blue shift in the optical functions when the values of the $\theta$-parameter are increased. These results can be useful for indirect experimental measures of the magnetoelectric polarizability which is proportional to the QED fine-structure constant $\alpha$.
\end{abstract}
%
\maketitle
\section{Introduction}\label{sec:intro}

Semiconductor quantum dots (QDs) have drawn the attention of low-dimensional physics in the last few decades due to the large number of possible applications in laser amplifiers\cite{mid-infrared,far-infrared}, photodetectors \cite{photodetector-IR,photodetector-UV}, high-speed electro-optical modulators \cite{modulators}, photonics \cite{Ramirez2019, Lujan2019},  and biosensing \cite{biosesing}. The current nanofabrication technologies allow to tune QD properties through the precise control in the particle size \cite{tunable-size,tunable-size2}, shape, \cite{tunable-shape}, composition,  and the number of electrons in such a dot \cite{tunable-number}. Especially, core/shell QDs (CSQD), usually formed by  semiconductor-insulator \cite{spherical-insulator} or semiconductor-semiconductor \cite{spherical-semiconductor,spherical-shell} materials, improve some properties compared with a QD's monocomponent.

Recently, thermal, magnetic, and optical properties due to electrons confined in several potential models have attracted considerable attention due to their significance in various scientific and technical fields \cite{parabolic-0,Castano2018}. The confining potential is decisive for the correct description of QD dynamics, as the energy levels splitting is largely determined by both the symmetry and the width of such potential. In that way, it has been established that a harmonic potential reproduces the main characteristics of such systems: for example, the effects caused by the electron-electron interaction in the energy spectrum \cite{parabolic-1,parabolic-2,parabolic-3} and its electronic structure \cite{parabolic-4}, and the effects of a topological defect on the thermo-magnetic and optical properties of QD \cite{castano2019}, among others~\cite{ConfiningPotential1,ConfiningPotential2,ConfiningPotential3}. On the other hand, the interaction of QDs with external magnetic fields is interesting in terms of the modifications on its physical properties such as the ground state energy and the addition of electron spectra~\cite{Magnetic1,Magnetic2}, the thermopower response~\cite{Magnetic3}, spin-orbit interactions~\cite{Magnetic4} and spin blockade~\cite{Magnetic5}. Moreover, several studies have been carried out regarding the modification of the nonlinear optical properties of QDs in a magnetized medium: Khordad studied the role of the spin-orbit interaction (SOI) and magnetic field in the electronic and optical properties of a double ring-shaped quantum dot~\cite{Opticalproperties1} as well as the impact of such coupling when the geometry is modified~\cite{Opticalproperties3,Opticalproperties4}; \c{C}akır \etal~have shown how the Zeeman coupling modifies the optical response of a spherical QD~\cite{Opticalproperties2}; Gul Kilic \etal~reported the influence of an hydrogenic impurity exposed to a magnetic field on the absorption coefficient and the refractive index changes~\cite{Opticalproperties5} and recently, Antil \etal~have combined the effect of pressure, temperature and magnetic fields in the optical response~\cite{Opticalproperties6}.

Due to the interesting electromagnetic response of QDs, in this paper we study the optical response of a core-shell QD when it is coupled to a topologic insulator. Topological Insulators (TIs) are unusual quantum materials admitting insulating bulk and conducting surface states protected by time-reversal-symmetry (TRS) \cite{Hasan,Qi Review}. When TRS is broken, several exotic phenomena are predicted to occur \cite{Hasan,Qi Review,Qi PRB} such as the quantum anomalous Hall (QAH) effect \cite{Qi PRB, Yu,Liu,Chang 1,Kou,Checkelsky,Chang 2,Liu-Wang,Kandala}, the quantized magneto-optical effect~\cite{Qi PRB,Tse,Maciejko,Okada,Wu,Dziom}, the topological magnetoelectric (TME) effect~\cite{Qi PRB,Wang,Morimoto,Essin,Mogi,Nomura}, the image magnetic monopole~\cite{Qi Science} and recently the reversed Vavilov-\v{C}erenkov radiation \cite{OJF-LFU-ORT-1}. The QAH and quantized magneto-optical effects have already been experimentally demonstrated in magnetic TI films~\cite{Chang 1,Kou,Checkelsky,Chang 2,Okada,Wu,Dziom,Mogi}.

The novel behavior of TIs was theoretically predicted by the pioneer works of Kane, Mele, and Bernevig~\cite{Kane-Mele
1,Kane-Mele 2, Bernevig-Hughes-Zhang}, and its experimental realization in two dimensional system was reported by Koenig \etal~\cite{Koenig et al}. A generalization in three dimensions the topological characterization of the quantum spin Hall insulator state can be found in Refs.~\cite{Fu-Kane-Mele, Moore-Balents, Roy}. Such behavior was predicted in several real materials, which included Bi${}_{1-x}$Sb$^{}_{x}$ as well as strained HgTe and $\alpha$-Sn~\cite{Fu-Kane}. Subsequently,  the experimental realization of the first 3D TI in Bi${}_{1-x}$Sb$^{}_{x}$ was reported by Hsieh \etal~\cite{Hsieh et al}. Later, the second generation of TIs, such as Bi${}_{2}$Se${}_{3}$, Bi${}_{2}$Te${}_{3}$ and Sb${}_{2}$Te${}_{3}$, were identified theoretically in Ref. \cite{Xia et al} and experimentally discovered in Refs. \cite{Xia et al, Zhang et al}. All these advances motivated the search of new TIs \cite{ANDO} and their classification into a periodic table where different classes of these materials can be identified~\cite{Hasan}. It is worth mentioning that a new type of TIs, called axion insulators (AXIs), has been recently proposed as a new arena to probe topological phases. They have the same bulk properties as 3D TIs, but the conducting surface states are protected by inversion symmetry, instead of TRS~\cite{AXIs}.

From the above, the coupling of material with a non-trivial magnetic response to semiconductor arrays can be interesting in terms of new physical phenomena and possible future applications (the magnetoresistance is a good example of such coupling~\cite{GMRFert,GMRGrunberg}), and therefore, the aim of the present work is to study the optical properties of a spherical 3D TI embedded in a single-electron system consisting of a spherical GaAs QD when the whole system is subjected to a uniform external magnetic field. 

This paper is organized as follows: in Sec.~\ref{Sec:Core-Shell} we describe the system under consideration. Sec.~\ref{Sec:ThetaED} is devoted to review the so-called $\theta$-electrodynamics and to present the Green's Function Method to obtain the induced electric and magnetic potentials. In Sec.~\ref{Sec_hamiltonian} we present the GaAs quantum Hamiltonian in the effective-mass approximation and the Lagrange-mesh method to solve the Schrödinger equation. Sec.~\ref{Sec_Opt_Prop} is dedicated to the intersubband optical properties and the discussion of the selection rules which give the allowed dipolar transitions. The results can be found in Sec.~\ref{Sec_results} and the conclusions in Sec.~\ref{Sec_Concl}.

\textit{Notation:} We denote the Minkowski metric with signature  $(+,-,-,-)$ by $\eta^\mu{}_\nu$  and we adopt the convention $\varepsilon^{0123}=1$ for the Levi-Civita symbol.


\section{Core-shell quantum dot}\label{Sec:Core-Shell}

\begin{figure}[h]
    \centering
    \includegraphics[scale=0.29]{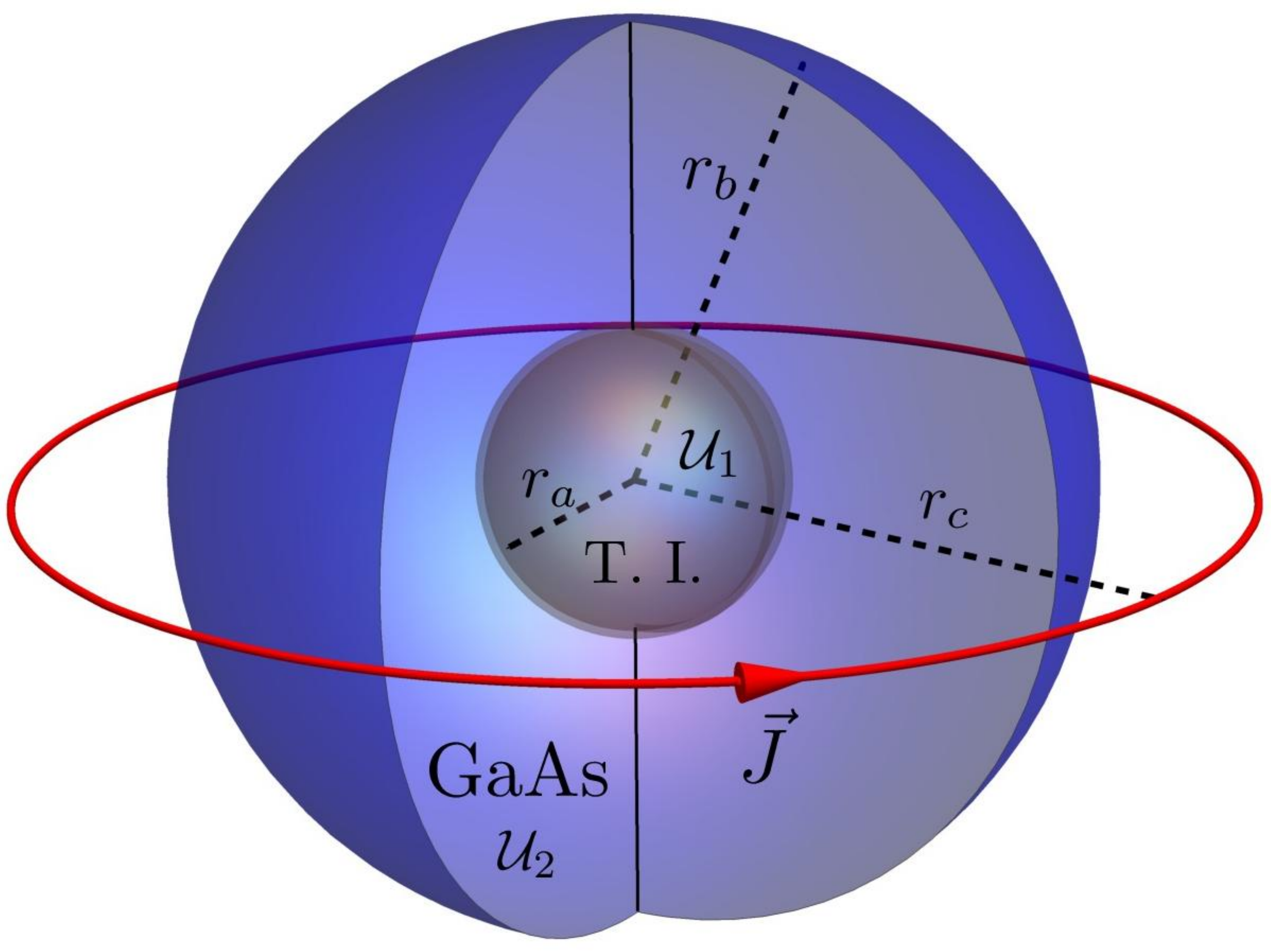}
    \caption{Scheme of the system: The topological insulator (TI) is a concentric sphere of radius $r_a$ inside of the spherical GaAs quantum dot of radius $r_b$. The system is placed in an external magnetic field in the $z$-direction induced by a circular current $\mathbf{J}$ with radius $r_c$. }
    \label{Esquema}
\end{figure}

 In the present model, we study a single electron confined in a 3D core-shell GaAs QD in the effective mass approximation. The inner region of the core-shell is made of a TI. The whole system is in the presence of an external and constant magnetic field, which is provided by an outer current loop. Figure~\ref{Esquema} schematizes the QD: the region $0<r<r_a$ is made of the TI and we suppose that it is inaccessible for the electron which can move only between $r_a<r<r_b$. The current loop has a radius $r_c\gg r_a,r_b$ and its current intensity $\mathbf{J}$ determines the magnitude of the magnetic field.
 
 The quantum Hamiltonian is constructed by computing the expressions for the electric and the magnetic potentials which interact with the free electron. Those fields are provided by the interaction between the TI and the external current, and they are obtained through the $\theta$-ED formalism.
 
 Regarding the manufacturing of the proposed system, the mismatch of the semiconductor-TI junction is an important point to take into account, particularly by its impact on the optical properties of core-shell QDs~\cite{Li-Chen}. It is well-known that Gallium Arsenide and other III-V semiconductors belong to the cubic system, while TIs are mainly hexagonal or trigonal~\cite{trigonal}, and both differ from their cell parameters\footnote{TlBiSe$_2$: Space group R-3m (166), $a=b=4.235\si{\angstrom}$ and $c=22.288\si{\angstrom}$~\cite{ICDD1}.\\
Bi$_2$Se$_3$: Space group R-3m (166), $a=b=4.1396\si{\angstrom}$ and $c=28.636\si{\angstrom}$~\cite{ICDD2}.\\
Bi$_2$Te$_3$: Space group R-3m (166), $a=b=4.385\si{\angstrom}$ and $c=30.483\si{\angstrom}$~\cite{ICDD3}.\\
GaAs: Space group F-43m (216), $a=b=c=5.6534\si{\angstrom}$~\cite{ICDD4}.}. This means that TIs grown over semiconductor substrates (or vice versa) exhibit a large lattice mismatch. The several surface defects may have impact in the electrons mobility and the optical properties when the so-called defect levels are created. Moreover, it is commonly reported that the difference in the cell parameters about 10$\%$ produces a high concentration of defects in the interface, in particular dislocations extending over thousand of Angstroms~\cite{librosolidstate} which implies a extended region with its own potential function, therefore, the description can be approximated as a multilayer system. Nonetheless, Xinyu Liu {\it et al.} grew by epitaxial methods Bi$_2$Te$_3$ and Bi$_2$Se$_3$ in a GaAs substrate~\cite{GaAsTI1,GaAsTI2}. Their results show a large mismatch  between Bi$_2$Te$_3$ and GaAs,  while Bi$_2$Se$_3$-GaAs interface is almost uniform, suggesting  that is possible to manufacture highly crystalline materials which is the case of our model. The discussion about the band junction and the potential choice is given later in Sec.~\ref{Sec_hamiltonian}.

\section{$\theta$-Electrodynamics}\label{Sec:ThetaED}
 Topological Insulators are best characterized as bulk magnetoelectrics with a quantized magnetoelectric response coefficient of the order of the fine-structure constant $\alpha=e^{2}/\hbar c$~\cite{Qi PRB, Essin}. Magnetoelectrics are materials wherein polarization can be created by an applied  magnetic, or a magnetization can be created by an applied electric field \cite{Fiebig}. If we denote by ${\mathbf E}$ the electric field and by ${\mathbf B}$ the magnetic field, TIs can be described as a special $\mathbf{E}\cdot\mathbf{B}$ magnetoelectrics whose topological field theory is a consequence of the addition of the term  ${\cal L}_\theta=-(\alpha/4\pi^2) \theta \, {\mathbf E}\cdot {\mathbf B}$ to the Maxwell Lagrangian  ${\cal L}_{\rm EM}$. In order to account for their electromagnetic response, the additional term can be rewritten as
\bea
{\cal L}_\theta=-\frac{\alpha\theta(\mathbf{x})}{16\pi^2}F_{\mu\nu}\T{F}^{\mu\nu},
\eea
and thus is identified with the Pontryagin term in 3+1 dimensions, where $F_{\mu\nu}$ is the electromagnetic tensor and $\T{F}_{\mu\nu}$ is the Hodge dual field strength electromagnetic tensor, defined as $\T{F}_{\mu\nu}=\frac{1}{2}\varepsilon^{\mu\nu\alpha\beta}F_{\alpha\beta}$~\cite{Urrutia1,Urrutia2,Liang,Huerta1,Huerta2}.

In the context of condensed matter, $\theta$ is known as the (scalar) magnetoelectric polarizability (MEP), but in field theory, the MEP is called the axion field, which generates axion electrodynamics and whose Lagrangian is just ${\cal L}_\theta$ plus the Maxwellian Lagrangian \cite{Wilczek}. However, we will consider $\theta$ as an additional parameter characterizing the material, in the same spirit as its permittivity $\epsilon$ and permeability $\mu$. In the axion theory the field $\theta$ is dynamical but here it will no longer have this characteristic. This forces us to restrict the name from axion electrodynamics to $\theta$-electrodynamics ($\theta$-ED) emphasizing that $\theta$ has no dynamics. Although ${\mathbf E}\cdot {\mathbf B}$ is a total derivative when it is coupled to the piecewise axion field $\theta$, it will contribute and provide considerable consequences at the surfaces and interfaces, where gradients of $\theta$ appear \cite{Essin}. The resulting Maxwell equations derived from ${\cal L}_\theta +{\cal L}_{\rm EM}$  are  those corresponding to  a standard material medium having  the modified   constitutive relations
\bea
 \mathbf{D}&=&\epsilon \mathbf{E}-({\alpha }\theta / \pi) \mathbf{B}\nn\\
 \mathbf{H}&=&\mathbf{B}+({\alpha }\theta / \pi) \mathbf{E}.
 \label{constitutiveecs}
\eea

Let us give some general comments on the values that $\theta$ can take when it is employed to model a TI. Retaining the values of $ \hbar$ and $c$, we recall that the electric and magnetic fields have dimensions of charge divided by distance squared in Gaussian units. In this way, the contribution of ${\cal L}_\theta$ to its corresponding action is $S_\theta=(c \hbar^2/e^4)\int dt \, d^3x \, {\cal L}_\theta $. Now, if we rewrite ${\mathbf E}\cdot {\mathbf B} $ in terms of the field strength tensor  $F_{\alpha\beta}=\partial_\alpha A_\beta-\partial_\beta A_\alpha$, where   $A_\mu$ is the electromagnetic potential, and considering a closed spacetime with no boundaries we get 
\begin{equation}
\frac{S_\theta}{\hbar}=\frac{\theta}{32 \pi^2 }\int d^4x\, \epsilon^{\alpha\beta\mu\nu} \frac{1}{e^2} F_{\alpha\beta} F_{\mu\nu}=\theta\,  C_2, \label{THETAACT}
\end{equation}
where $C_2$ is an integer.  This is because in such spaces the dimensionless integral in Eq. (\ref{THETAACT}) is equal to $32 \pi^2\, C_2$, where $C_2$  is the second Chern number of the manifold \cite{FUJIKAWA}. Under changes of $\theta$, the quantity $\exp(-i S_\theta/\hbar)$ must remain invariant, which means that two values of $\theta$ differing by an integer  multiple of $2 \pi$ are equivalent. Furthermore imposing TRS  yields to new constraints  on the values of $\theta$. Since $ \mathbf{E}\cdot \mathbf{B}$ is odd under TRS, one could think that the only allowed value would be $\theta=0$ (modulo  $2 \pi)$. Nevertheless, the condition $\exp(-iS_\theta/\hbar)= \exp(+iS_\theta/\hbar)$ yields to the possibility of having $\theta= \pi $. In such a way that we obtain two families of magnetoelectric materials described by the choices $\theta_1= 0$ (normal insulators) and $\theta_2= \pi$ (TIs). Both values of $\theta$ are defined modulo  $2 \pi$.

As we mentioned above, the integrand $\mathbf{E}\cdot \mathbf{B}$ of $S_\theta$ in Eq.~(\ref{THETAACT}) is a total derivative which is directly verified by the Bianchi identity, so new contributions will only come from gradients of $\theta$. Thus, when at the interface $\Sigma$ two materials have different constant values of $\theta$, then modifications to Maxwell equations will arise because $\partial_\mu\theta\neq0$ holds. In this case the action of Eq.~(\ref{THETAACT}) can be integrated yielding a $2+1$ action at the boundaries  corresponding to the  Chern-Simons term. This means that at the boundaries of a 3D TI we have a QAH effect  associated to each value of $\theta$ with Hall conductivity given by $\sigma_{{\rm H}}=\theta e^2/2 \pi h $. In this way, the contribution  to the  total Hall conductivity from the interface $\Sigma$ between  a TI and a regular insulator  is

\begin{equation}
\sigma_H^{\Sigma}=\frac{e^2}{h}\left(\frac{1}{2}+ \tilde{m}\right),
\end{equation} 
since two values of $\theta$ differing by an integer multiple $\tilde{m}$ of $2\pi$ are  equivalent.

The half integer contribution to $\sigma^\Sigma_H$ is a bulk property, which allows us to distinguish this case from that of a 2D surface gapped crystal having $\sigma_H= N e^2/h$, with $N$ an integer, thus showing that both conditions  are not topologically equivalent. When dealing with a TRS invariant TI  in a region with no boundaries, the number $m$ remains undetermined. The integer part of $\sigma^\Sigma_H$
becomes resolved   only in the presence of a boundary between two TIs with different values of $\theta$, when  TRS is broken  by  gapping the interface. So we provide an adiabatic transition between those two topologically inequivalent insulators and the value of $\tilde{m}$  depends on  the specific properties of such breaking. Such TRS breaking is usually realized  by an external magnetic field across the interface  or by  a  magnetic doping of the surface.

To describe the interaction between a 3D topologically insulating ponderable media with electromagnetic external sources, we must characterize their electromagnetic response by their dielectric permittivities $\epsilon_1$, $\epsilon_2$ and their MEP $\theta_1$, $\theta_2$. As a first step in dealing with optical properties, we will consider the permittivities and MEPs frequency independent. For definiteness, we deal with materials having $\mu=1$. Let us consider two finite spherical and concentric magnetoelectric media separated by a spherical interface $\Sigma$ located at $r_a$ inside of a finite sphere of radius $r_b$. The inner region $\mathcal{U}_1$ ($0<r<r_a$) is filled with a TI, and the outer region $\mathcal{U}_2$ ($r_a<r<r_b$) will contain GaAs, as shown in Fig. \ref{Esquema}. Additionally, we assume the field $\theta$ as piecewise constant taking the values $\theta=\theta_1$ in $\mathcal{U}_1$ and $\theta=\theta_2=0$ in $\mathcal{U}_2$ (because GaAs is not a TI). This is expressed as
\bea
\theta(r)=\Theta(r_a-r)\theta_1,
\eea
where $\Theta(r_a-r)$ is the Heaviside function with $\Theta(r_a-r)=1,$ for $r_a>r$, otherwise $\Theta(r_a-r)=0$. The whole dielectric permittivity of the system is piecewise constant and is described as
\bea
\epsilon(r)=\Theta(r_a-r)\epsilon_1+\Theta(r-r_a)\epsilon_2.
\eea
In this particular case, we can take the action for the effective field theory describing the electromagnetic response of this media as
\begin{eqnarray}\label{Action}
S[\Phi ,\mathbf{A}]&=&\int_\mathcal{M} dt\,d^{3}\mathbf{x}\left[ \frac{1}{8\pi }\left(
\epsilon \mathbf{E}^{2}-\mathbf{B}^{2}\right)\right.\nn\\
&&\left.-\frac{\alpha }{4\pi ^{2}}\theta(r)\,\mathbf{E}\cdot \mathbf{B}-\varrho \Phi +\mathbf{J}\cdot \mathbf{A}\right]\;,
\end{eqnarray}
where the integral runs over a (3+1)-dimensional spacetime $\mathcal{M}=\mathcal{U}_1\cup\mathcal{U}_2\times\mathbb{R}$, with $\mathbb{R}$ as the temporal axis. Also in Eq. (\ref{Action}), $\varrho$ and $\mathbf{J}$ are external charge and current densities, as usual $\alpha$ is the fine-structure constant and for the moment everything is in Gaussian units with $c=\hbar=1$.

Let us emphasize two important points. The first one is the key role played by the $\theta$-term in (\ref{Action}), which supplies the magnetoelectric contribution representing the TME effect, a fundamental feature of TIs. Secondly, we are describing each medium by a constant MEP $\theta$ in the bulk with $\theta=0$ for a normal insulator and $\theta=\pi$ for a TI, whose value has a gauge-invariant and topological origin.

It can be shown that the homogeneous Maxwell equations related to Eq. (\ref{Action}) are the usual ones,
\bea
\nabla \cdot \mathbf{B}=0,\quad\nabla \times \mathbf{E}=-\frac{\partial \mathbf{B}}{\partial t}, \label{HOMEQ}
\eea
which enables us to relate the electromagnetic fields $\mathbf{E}$ and $\mathbf{B}$ with the electromagnetic potentials $\Phi$ and $\mathbf{A}$ in the standard form
\bea
\mathbf{E}=-\frac{\partial\mathbf{A}}{\partial t}-\nabla\Phi ,\quad \mathbf{B}=\nabla \times \mathbf{A}.\label{Potenciales}
\eea

As discussed previously $\mathbf{E}\cdot\mathbf{B}$ is a total derivative, which means that the dynamics in the bulk is the same as in standard ED. So, all the new effects induced by $\mathcal{L}_\theta$ arise on the interface and manifest themselves as a consequence of the modified boundary conditions there. After performing the variation of the action in Eq. (\ref{Action}), we obtain the following set of modified Maxwell equations:
\begin{eqnarray}
\nabla \cdot\left[\epsilon(r)\mathbf{E}\right]&=&4\pi \varrho +\tilde{\theta}\delta (r-r_a)\mathbf{B}\cdot \mathbf{\hat{e}}_u\;,  \label{Gauss E} \\
\nabla \times \mathbf{B}-\epsilon(r)\frac{\partial \mathbf{E}}{\partial t}
&=&4\pi\mathbf{J}+\tilde{\theta}\delta (r-r_a)\mathbf{E}\times \mathbf{\hat{e}}_u\;,  \label{Ampere}
\end{eqnarray}
where $\mathbf{\hat{e}}_u$ is the outward unit vector normal to the interface $\Sigma$ located at $r=r_a$ and 
\begin{equation}
{\tilde{\theta}=\alpha\theta/\pi.}
\label{TILDE_THETA}
\end{equation}
In the case of a TI located in region $\mathcal{U}_1$  of Fig. \ref{Esquema} ($\theta_1=\pi$) in front of a regular insulator ($\theta_2=0$) in region $\mathcal{U}_2$,  we have \begin{equation}
{\tilde \theta}=\alpha(2\tilde{m}+1), \label{TILDE_THETA_1}
\end{equation}
where $\tilde{m}$ is an integer depending on the details of the TRS breaking at the interface. The main feature of the
above equations is that they introduce additional field-dependent effective charge and current densities
\begin{equation}
\varrho _{\theta }=\frac{1}{4\pi }\tilde{\theta}\delta (r-r_a)\mathbf{B}\cdot 
\mathbf{\hat{e}}_u,\qquad \mathbf{J}_{\theta }=\frac{1}{4\pi }\tilde{\theta}\delta (r-r_a)\mathbf{E}\times \mathbf{\hat{e}}_u,  \label{EFFSOURCES}
\end{equation}
with support only on the interface $\Sigma$  between the two media. Consequently, the standard Maxwell equations remain valid in the bulk. Let us remind that although we are modeling the electromagnetic response of the TI by an effective theory, the TI fermionic degrees of freedom are integrated, in particular the ones related to the TI surface states~\cite{Fermionic integration}. These ones are codified in Eqs. (\ref{EFFSOURCES}) and their effects will propagate through the bulk.

On the other hand, it is shown that the densities of Eq.~(\ref{EFFSOURCES}) satisfy the conservation equation 
\begin{equation}
\nabla\cdot\mathbf{J}_\theta + \frac{\partial \varrho_\theta}{\partial t}=0,
\end{equation}
which can be readily verified  by  using Faraday's law together with the relation 
\begin{equation}
\nabla\delta (r-r_a)\cdot\left(\mathbf{E}\times\mathbf{\hat{e}}_u\right)=\left[ \partial _{r}\delta (r-r_a)\right]\mathbf{\hat{e}}_r\cdot\left(\mathbf{E}\times\mathbf{\hat{e}}_u\right)=0.
\end{equation}
Here we remark that Eqs.~(\ref{Gauss E}) and (\ref{Ampere}) can also be obtained starting  from the standard Maxwell equations in a  material medium \cite{Schwinger,Jackson}:
\begin{eqnarray}
\nabla \cdot \mathbf{D}=4\pi \varrho &,&\;\nabla \times \mathbf{H}=\frac{\partial \mathbf{D}}{\partial t}+4\pi \mathbf{J}\;, \\
\nabla \cdot \mathbf{B}=0 &,&\;\;\;\nabla \times \mathbf{E}=-\frac{\partial\mathbf{B}}{\partial t}\;,
\end{eqnarray}
together with the constitutive relations of Eqs.~(\ref{constitutiveecs}).

Assuming that the time derivatives of the fields are finite in the vicinity of the interface $\Sigma$, the modified Maxwell equations (\ref{Gauss E}) and (\ref{Ampere}) yield the following boundary conditions (BCs) 
\begin{equation}
\left[\epsilon\mathbf{E}_{r}\right]_{r=r_a^{-}}^{r=r_a^{+}} =\tilde{\theta}\mathbf{B}_{r}|_{r=r_a} 
,\left[\mathbf{\hat{e}}_u\times\mathbf{B}\right]
_{r=r_a^{-}}^{r=r_a^{+}}=-\tilde{\theta}\mathbf{\hat{e}}_u\times\mathbf{E}|_{r=r_a},\label{Conditions 1}
\end{equation}
\begin{equation}\label{Conditions 2}
\left[\mathbf{B}_{r}\right] _{r=r_a^{-}}^{r=r_a^{+}}=\mathbf{0},\quad\left[ \mathbf{\hat{e}}_u\times\mathbf{E}\right]_{r=r_a^{-}}^{r=r_a^{+}}=\mathbf{0}\;,
\end{equation}
for vanishing external sources at $r=r_a$. These BCs are derived
either by integrating the field equations over a spherical shell across the interface or by using the Stokes theorem over a closed circuit crossing the interface. The notation is $\left[ \mathbf{V}\right] _{r=r_a^{-}}^{r=r_a^{+}}=\mathbf{V}%
(r=r_a^{+})-\mathbf{V}(r=r_a^{-})$, $\mathbf{V}\big|_{r=r_a}=\mathbf{V}(r=r_a)$,
where $r=r_a^{\pm }$ indicates the limits $r=r_a\pm \eta $, with$\;\eta\rightarrow 0$,$\;$ respectively. The continuity conditions of Eqs.~(\ref{Conditions 2}) imply that the right-hand sides of the discontinuity conditions of Eqs.~(\ref{Conditions 1}) are well defined and they represent self-induced surface charge and surface current densities, respectively. Those BCs again clearly illuminate the ME effect, which is localized just at the interface $\Sigma$  between the two spherical media where the effective charge and currents densities of Eq.~(\ref{EFFSOURCES}) lie. As we shall see at the end of the current section, the impact of these surface contributions will manifest in the electric and magnetic field.

\subsection{Green's Function Method}\label{sec:GF Method}

In this subsection we adapt the Green's function method from Ref. \cite{Urrutia3} to obtain the modified electric and magnetic fields due to the presence of the TI embedded in another magnetoelectric medium when both of them are endowed with dielectric properties. By knowing the GF of a certain configuration one is able to compute the electromagnetic fields for an arbitrary distribution of sources, as well as to solve  problems with given Dirichlet, Neumann or Robin boundary conditions on surfaces of arbitrary geometry. For our purposes, we need to restrict ourselves to contributions of free external and time-independent sources $J^\mu=(\varrho,\mathbf{j})$ located outside the interface $\Sigma$. After inserting the potentials of Eq.~(\ref{Potenciales}) into the inhomogeneous Maxwell equations~(\ref{Gauss E}) and~(\ref{Ampere}), we find the following coupled equations 
\bea
\left[\mathcal{O}^\mu_{\;\;\nu}\right]_\n{x}A^\nu(\n{x})=4\pi J^\mu,\label{ecforA}
\eea
where the Coulomb gauge $\nabla\cdot\mathbf{A}=0$ was assumed and the differential operator $\left[\mathcal{O}^\mu_{\;\;\nu}\right]_\n{x}$ explicitly reads as
\bea
\left[\mathcal{O}^\mu_{\;\;\sigma}\right]_\n{x}=\left( 
\begin{array}{cccc}
\mathcal{O}_r^{(\epsilon)} &\hat{\Gamma}_{x} &\hat{\Gamma}_{y} & \hat{\Gamma}_{z} \\ 
\hat{\Gamma}_{x} & \mathcal{O}_r^{(1)} & 0 & 0 \\ 
\hat{\Gamma}_{y} & 0 & \mathcal{O}_r^{(1)} & 0 \\ 
\hat{\Gamma}_{z} & 0 & 0 & \mathcal{O}_r^{(1)}%
\end{array}
\right)\;,
\eea
with $\hat{\Gamma}_{k}=i\T{\theta}\delta(r-r_a)\hat{L}_{k}/r$, $\hat{L}_{k}=-i\left(\mathbf{x}\times\nabla\right)_{k}$ are the components of the angular momentum ($k=x,y,z$), $\mathcal{O}_r^{(1)}=-\nabla^2$ and
\bea
\mathcal{O}_r^{(\epsilon)}=\epsilon(r)\nabla^{2}-\frac{\partial\epsilon(r)}{\partial r}\frac{\partial}{\partial r}.
\eea

The BCs from Eqs.~(\ref{Conditions 1}) and~(\ref{Conditions 2}) now reduce to
\bea
A^{\mu }(\n{x})\big|_{r=r_a^{-}}^{r=r_a^{+}}&=&0\nn\\
\left[\epsilon(r)\partial_{r}A^{0}\right]_{r=r_a^{-}}^{r=r_a^{+}}&=&-\tilde{\theta}\mathbf{\hat{e}}_r\cdot\left(\nabla \times \mathbf{A}\right)\big|_{r=r_a},\nonumber\\
\left[\mathbf{\hat{e}}_u\times\left(\nabla \times \mathbf{A}\right)\right]\big|_{r=r_a^{-}}^{r=r_a^{+}}&=&\tilde{\theta}\mathbf{\hat{e}}_u\times\nabla\times A^{0}\big|_{r=r_a}.
\label{BCFG}
\eea

Next we introduce the GF ${G^\sigma}_\nu(\n{x},\n{x}')$ which satisfies
\bea
\left[\mathcal{O}^\mu_{\;\;\sigma}\right]_\n{x}{G^\sigma}_\nu(\n{x},\n{x}')=4\pi\nmn\delta^{(3)}(\mathbf{x}-\mathbf{x}'),
\label{ecforG}
\eea
together with the BCs arising from Eq.~(\ref{BCFG}) and the following BCs that our system imposes, i.e.
\bea
A^\mu(\n{x})|_{r=0}<\infty,\quad A^\mu(\n{x})|_{r=r_b}<\infty,
\label{Conditions 3}
\eea
the 4-potential is given by
\bea
A^\mu(\n{x})=\int_{V'}d^3\n{x}'{G^\mu}_\nu(\n{x},\n{x}')J^\nu(\n{x}'),\label{A and GF}
\eea
determined up to homogeneous solutions of Eq.~(\ref{ecforA}).

The solution of Eq. (\ref{ecforG}) for spherical geometry with a nonmagnetic TI and with the same discontinuity across the surface $r=r_a$ for $\theta(r)$ and $\epsilon(r)$ was found in Ref.\cite{Urrutia3} by the GF method. However this solution works only for the standard boundary conditions at infinity in contrast with Eq.~(\ref{Conditions 3}), therefore it is necessary to modify such solution in order to satisfy our BCs. First, we will provide a brief review of the procedure to solve Eq. (\ref{ecforG}) and then we will show how to correctly impose BCs. The full construction of the GF can be found in Appendix B of Ref.~\cite{Urrutia3}.
\begin{figure*}
    \centering
    \includegraphics[scale=0.45]{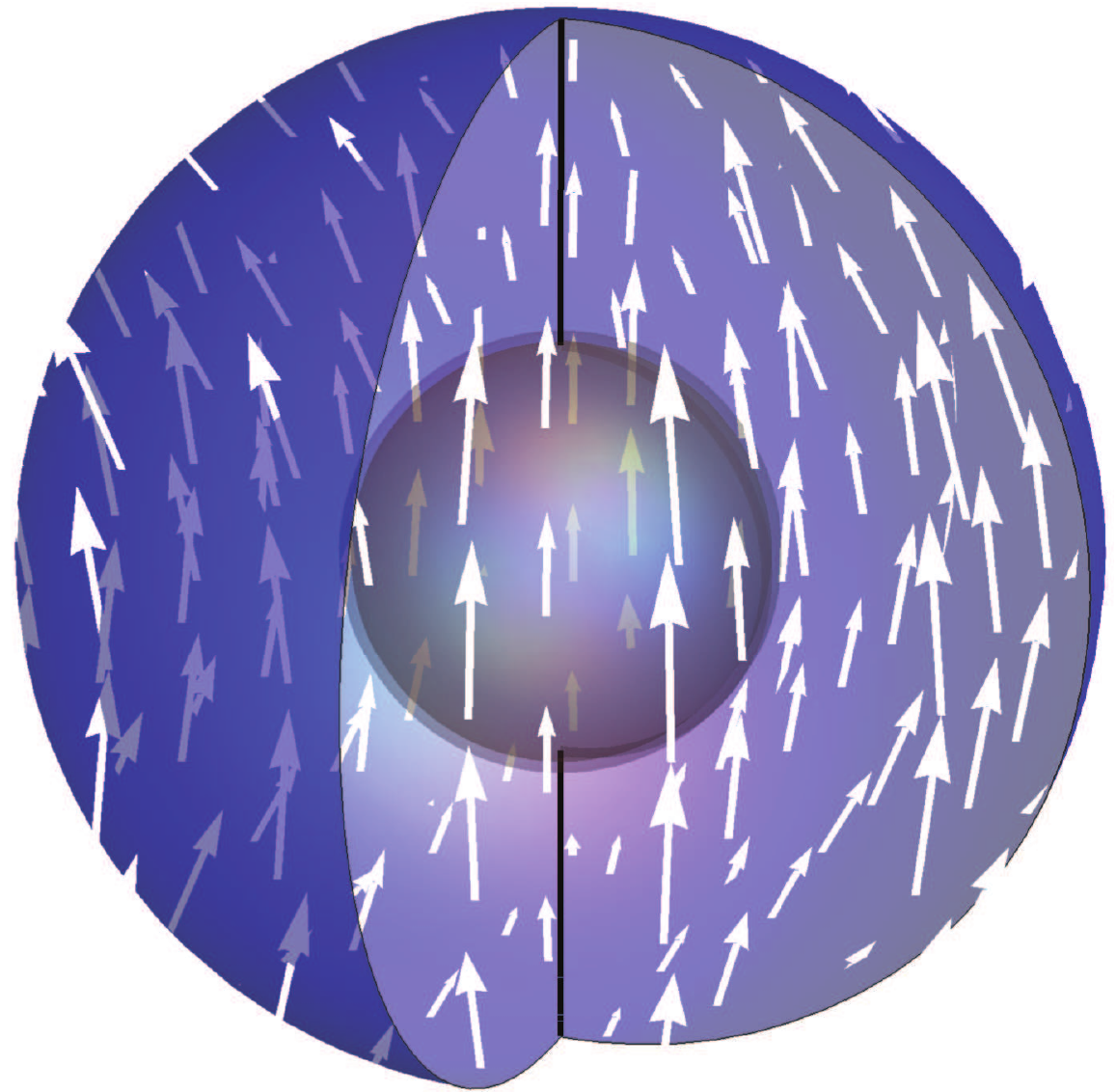} \hspace{1cm}\includegraphics[scale=0.45]{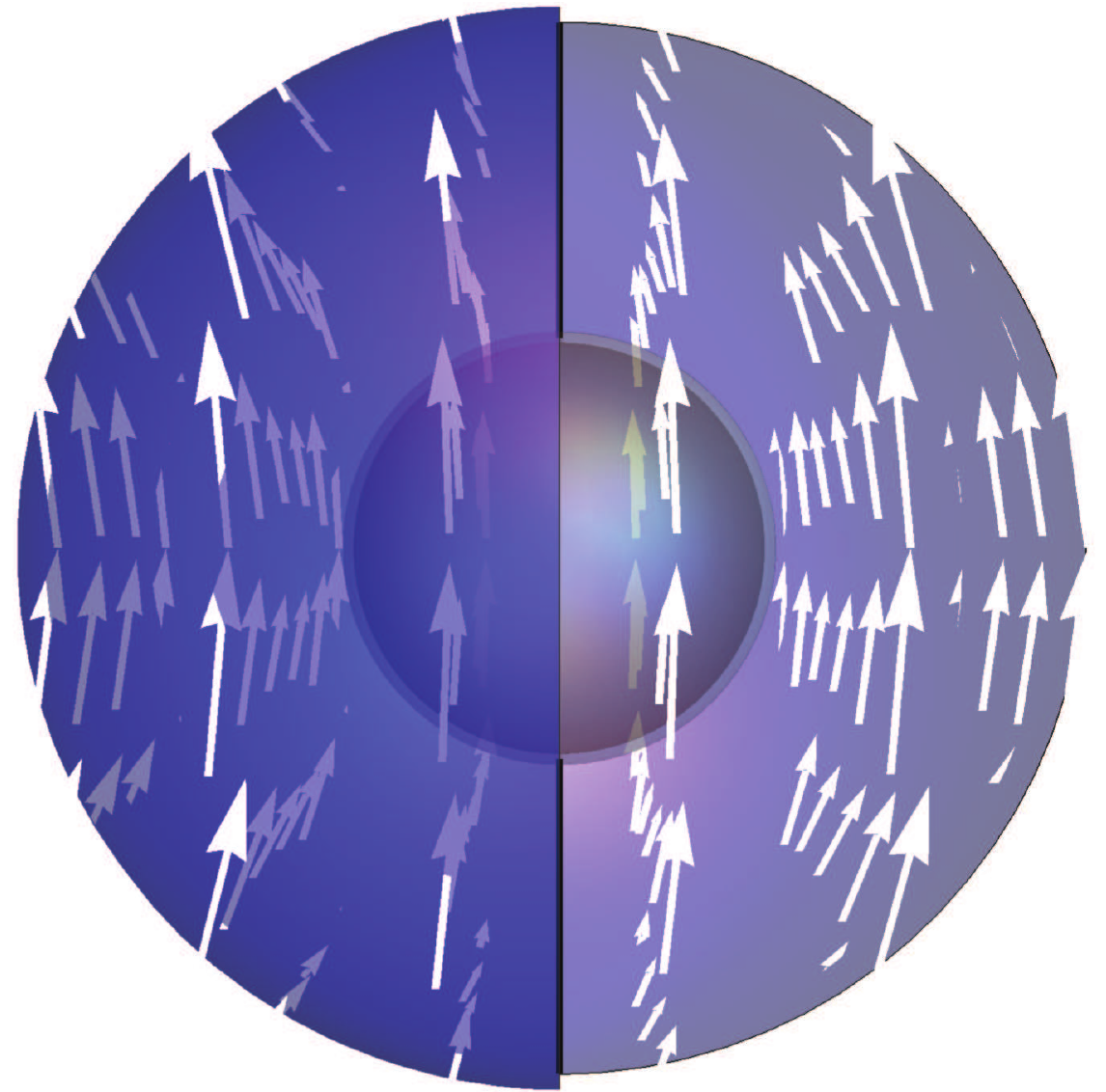}\hspace{1cm}\includegraphics[scale=0.45]{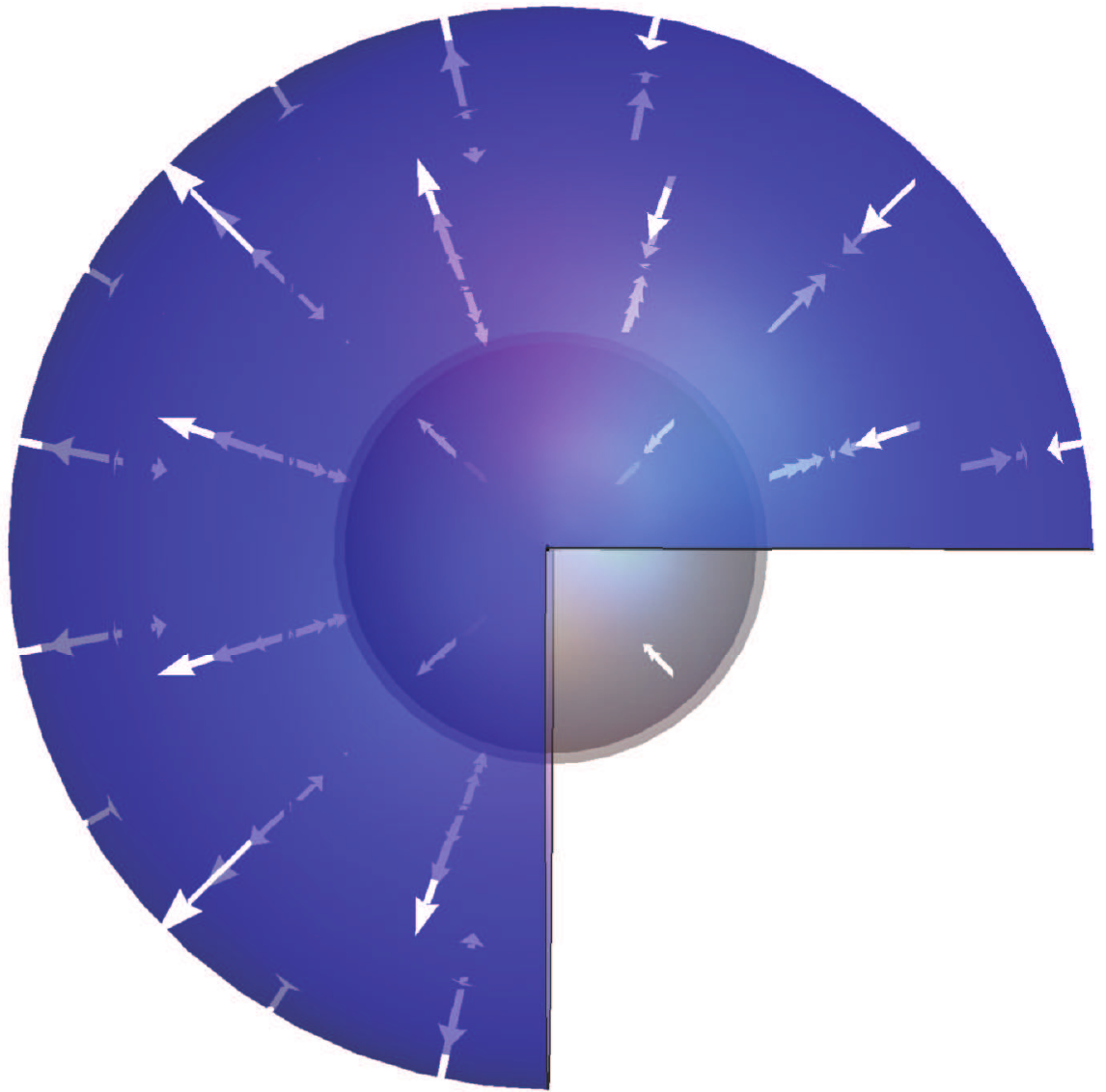}
    \caption{Electric field from Eq.~(\ref{ElectricField}) normalized by the factor $\tilde{\theta}Br_a^2/c\mu_0\epsilon_1$ for a QD with $r_a=2$nm, $r_b=5$nm and $r_c=200$ nm. From left to the right: Electric field in the full space $(x,y,z)$, in the $xz$-plane and in the $xy$-plane.}
    \label{Esquema2}
\end{figure*}

By using spherical coordinates ($r,\vartheta,\phi$) and since the square of angular momentum commutes with $\left[\mathcal{O}^\mu_{\;\;\sigma}\right]_\n{x}$, the GF can be written as
\bea
&&{G^\mu}_\nu(\n{x},\n{x}')\nn\\
&=&4\pi\sum_{l=0}^\infty\sum_{m=-l}^{+l}\sum_{m'=-l}^{+l}g^\mu_{lmm';\nu}(r,r')Y_{lm}(\vartheta,\phi)Y_{lm}^*(\vartheta',\phi'),\nn\label{GFM}\\
\eea
where $g^\mu_{lmm';\nu}(r,r')$ is the reduced GF, which satifies the equation
\bea
\sum_{m''=-l}^{+l}\hat{\mathcal{R}}^{\mu}_{lmm'';\sigma}g^\sigma_{lm''m';\nu}(r,r')=\eta^{\mu}_{\;\;\nu}\delta_{mm'}\frac{\delta(r-r')}{r^2}\;,
\eea
with $\hat{\mathcal{R}}^{\mu}_{lmm'';\sigma}=\langle lm|\left[\mathcal{O}^\mu_{\;\;\sigma}\right]_\n{x}|lm''\rangle$. The calculation of Ref.~\cite{Urrutia3} shows that the solution for the various components of the reduced GF is
\bea
&g^0_{lmm';0}(r,r')&=\delta_{mm'}\mathcal{F}_l^{(\epsilon)}(r,r')\nn\\
&&-\delta_{mm'}r_a^2\T{\theta}^2l(l+1)\mathcal{F}_l^{(1)}(r_a,r_a)S_l^{(\epsilon,\epsilon)}(r,r'),\nn\\
&g^i_{lmm';0}(r,r')&=-ir_a\T{\theta}\langle lm|\hat{L}^{i}|lm'\rangle S_l^{(1,\epsilon)}(r,r')\;,\label{red GFM}\\
&g^0_{lmm';i}(r,r')&=ir_a\T{\theta}\langle lm|\hat{L}_{i}|lm'\rangle S_l^{(\epsilon,1)}(r,r')\;,\nn\\
&g^i_{lmm';j}(r,r')&=\eta^i_{\;\;j}\delta_{mm'}\mathcal{F}_l^{(1)}(r,r')\nn\\
&&+r_a^2\T{\theta}^2\langle lm|\hat{L}^{i}\hat{L}_{j}|lm'\rangle
\mathcal{F}_l^{(\epsilon)}(r_a,r_a)S_l^{(\epsilon,1)}(r,r'),\nn
\eea
where
\bea
S_l^{(u,v)}(r,r')=\frac{\mathcal{F}_l^{(u)}(r,r_a)\mathcal{F}^{(v)}_l(r_a,r')}{1+r_a^2\T{\theta}^{2}l(l+1)\mathcal{F}^{(1)}_l(r_a,r_a)\mathcal{F}^{(\epsilon)}_l(r_a,r_a)}\label{S r r'}\nn\\
\eea
with $u,v=1,\epsilon$. Here $\mathcal{F}^{(u)}_l(r,r')$ are the free reduced GFs in absence of the spherical TI and should have the appropriate BCs given in Eq.~(\ref{Conditions 3}). The imposition of both requirements is achieved by taking the free reduced GFs that describe the electrodynamics inside a dielectric and nonmagnetic spherical shell of radius $r_b$. Thus, we have to use
\begin{equation}\label{free GF 1}
\mathcal{F}^{(\epsilon)}_{l}(r,r')=\frac{r^{l}_{<}}{\epsilon_1}\left[\frac{1}{r^{l+1}_{>}}+\frac{(\epsilon_1-\epsilon_2)(l+1)}{(\epsilon_1+\epsilon_2)l+1}\frac{r^{l}r'^{l}}{r_b^{2l+1}}\right]\;,
\end{equation}
which is the free reduced GF of the dielectric spherical shell interior and 
\begin{equation}\label{free GF 2}
\mathcal{F}^{(1)}_{l}(r,r')=\frac{r^{l}_{<}}{r^{l+1}_{>}}-\frac{r^{l}r'^{l}}{r_b^{2l+1}},
\end{equation}
the free reduced GF of the inside of a conducting spherical shell~\cite{Schwinger}. The notation $r_{<}(r_{>})$ represents the smaller (larger) of $r$ and $r'$.\\

These free reduced GFs solve the equations
\bea
\mathcal{O}_r^{(\epsilon)}\mathcal{F}^{(\epsilon)}_l(r,r')&=&\frac{\delta(r-r')}{r^2}\;,\\
\mathcal{O}_r^{(1)}\mathcal{F}^{(\epsilon)}_l(r,r')&=&\frac{\delta(r-r')}{r^2}\;.
\eea

\subsection{The magnetic field and the induced electric field}\label{Fields}
For our specific problem, the external current source is
\bea
\varrho(\n{x}')&=&0,\nn\\
\n{j}(\n{x}')&=&\frac{\mathcal{I}}{r_c}\sin\vartheta'\delta(r'-r_c)\left(\sin\phi'\hat{\mathbf{e}}_x-\cos\phi'\hat{\mathbf{e}}_y\right),\label{j}
\eea
where $\mathcal{I}$ is the current and the condition $r_c\gg r_a$ is imposed to ensure that the external magnetic field is oriented, uniform and constant in the direction of $\hat{\mathbf{e}}_z$.

After convoluting through Eq. (\ref{A and GF})
the GF (\ref{GFM}) with the charge and current densities of Eqs.~(\ref{j}), we obtain the following electric and magnetic potentials:
\begin{equation}
\Phi(\n{x})=\frac{\tilde{\theta}Br_a^3}{c\mu_0\epsilon_1}\left[\frac{1}{r^2}+2\frac{\epsilon_1-\epsilon_2}{\epsilon_1+\epsilon_2+\epsilon_0}\frac{r}{r_b^3}\right]\cos\vartheta\;.
\label{Phi SI}
\end{equation}
and 
\bea
\mathbf{A}(\mathbf{x})=\frac{B r}{2}\sin\vartheta\,\mathbf{\hat{e}}_\phi,
\label{vec_pot}
\eea
where we identified
\begin{equation}\label{B0}
B=\frac{3\mu_0\mathcal{I}}{2r_b^{3}r_c}(r_c^{3}-r_b^{3}),
\end{equation}
and both potentials are already expressed in SI units with $c=1/\sqrt{\epsilon_0\mu_0}$. The detailed calculation of the 4-potential can be found in Appendix~\ref{CalculoAmu}. Henceforth, $r$ will be restricted to region $\mathcal{U}_2$ (recall Fig.~\ref{Esquema}).

For the sake of completeness, we compute the electric and magnetic fields generated by the presence of the TI, which take the form
\bea
\mathbf{E}(\mathbf{x})&=&-\frac{\tilde{\theta}Br_a^3}{c\mu_0\epsilon_1r^3}\left(2\cos\vartheta\,\mathbf{\hat{e}}_r+\sin\vartheta\,\mathbf{\hat{e}}_\vartheta\right)\nn\\
&&+\frac{2\tilde{\theta}Br_a^3}{c\mu_0\epsilon_1r^3_b}\frac{\epsilon_1-\epsilon_2}{\epsilon_1+\epsilon_2+\epsilon_0}\,\mathbf{\hat{e}}_z,
\label{ElectricField}
\eea
and
\bea
\mathbf{B}(\mathbf{x})&=&B\,\mathbf{\hat{e}}_z.\label{MagneticField}
\eea

Finally, we would like to emphasize that the TME in this static problem plays a crucial role, because the current $\mathcal{I}$, apart from generating the vector potential $\mathbf{A}(\mathbf{x})$ and therefore the magnetic field $\mathbf{B}(\mathbf{x})$, also provides the scalar potential $\Phi(\mathbf{x})$ and consequently the electric field $\mathbf{E}(\mathbf{x})$, which explicitly depends on $\tilde{\theta}$ and $B$ as one can appreciate in Eqs. (\ref{Phi SI}) and (\ref{ElectricField}). The nontrivial spatial behavior of the electric field is illustrated in Fig. \ref{Esquema2}.

\section{Hamiltonian and Lagrange-mesh  Method}
\label{Sec_hamiltonian}
From the expressions for the vector and scalar potential, see Eqs. (\ref{Phi SI}) and (\ref{vec_pot}), the quantum Hamiltonian for a spinless electron in the effective-mass approximation moving the region $\mathcal{U}_2$ (GaAs), with the minimal coupling prescription can be written as
\bea
\hat{H}&=&\frac{1}{2m^*}\left(\n{p}-q\n{A}\right)^2+\hat{V}(r)+q\,\Phi(\n{x}),
\eea
where $q$ denotes the charge of the electron  and $m^*$ is the effective electron mass, taken as a constant in order to neglect the effects produced by the conduction electrons which are disregarded in the present work \cite{conductionband1,conductionband2,conductionband3}.

\begin{figure}[h]
    \centering
    \includegraphics[scale=0.75]{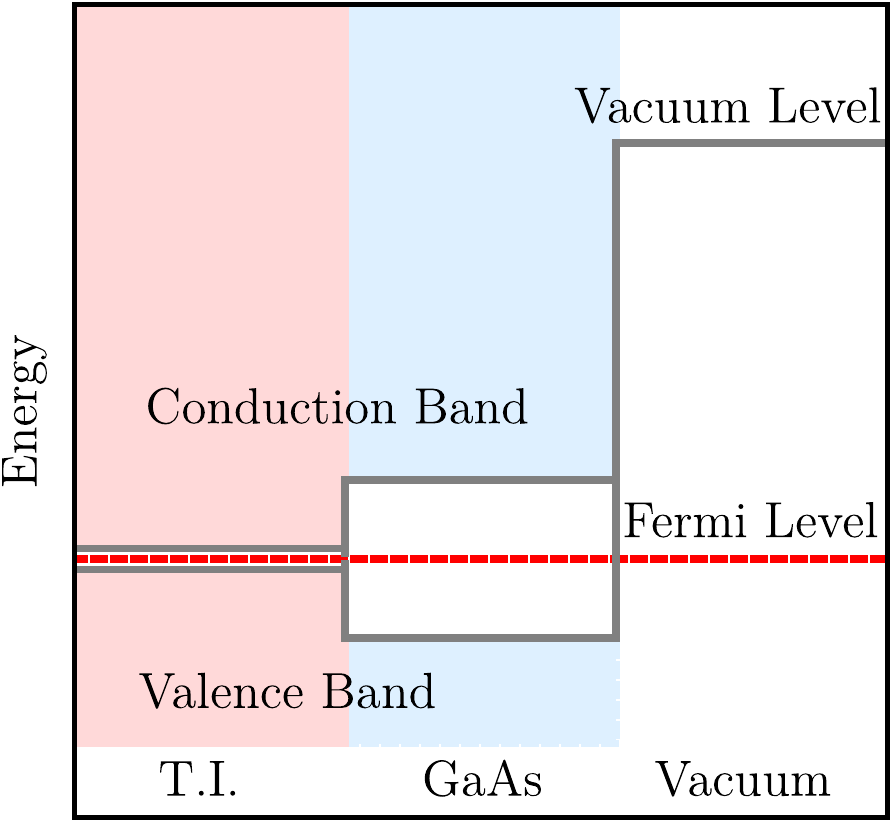}
    \caption{Sketch of the energy band junctures for the system TI-GaAs-Vacuum which can be classified as a core/shell QD type I~\cite{Energies}.}
    \label{Fig:Junction_Sketch}
\end{figure}

To simplify our calculations, we assume the hard-core approximation, i.e., the electrons can only move in the GaAs region. Such assumption can be justified as a first approximation of the real scenario, because topological insulators appear with band gaps smaller than common semiconductors, e.g., 1.52 eV for GaAs~\cite{DuquePhysB} (semiconductor), 0.105 eV for Bi$_2$Te$_3$, 0.090 eV for Sb$_2$Te$_3$~\cite{Shvets}, and 0.35 eV for TiBiSe$_2$~\cite{YLChen_et_al,Sato} (TIs). Therefore, according with Ref.~\cite{Energies} our core-shell can be classified as a Type I exemplified in Fig.~\ref{Fig:Junction_Sketch}, which shows a sketch of the band TI-GaAs-Vacuum juncture constructed by equating the Fermi energy of each material. Keeping this in mind and ignoring the conduction band charge carriers ($m^*$ is constant), from Fig.~\ref{Fig:Junction_Sketch} the lowest valence energy state lies in the GaAs region. Thus, the electrons prefer to move on the shell material. A more detailed systematic study without the hard-core prescription, which will take into account the interaction with the TI surface states \cite{TInanoparticle}, and the dynamics inside the core is in preparation and will soon be reported.

In this way, the term $\hat{V}(r)$ corresponding to the confining potential, which is taken as parabolic inside of the GaAs, is written as:
\bea
\label{confining_potential}
\hat{V}(r)=\left\{\begin{matrix}
\frac{1}{2}m^*\omega^2r^2, &\text{for}\;\;\; r_a\leq r\leq r_b\\ 
\infty,&\text{otherwise}&
\end{matrix}\right.,
\eea
where
\bea
\omega^2=\frac{V_0}{m^*(r_a-r_b)^2}\;.
\eea

In coordinate representation and recalling Eqs.~(\ref{Phi SI}) and (\ref{vec_pot}), the Hamiltonian reads
\bea
\label{Hamiltonian}
\hat{H}&=&-\frac{\hbar^2}{2m^*}\nabla^2+\frac{m^*}{2}\left(\omega^2+\frac{\omega_c^{2}}{4}\sin^2\vartheta\right)r^2+\frac{i\hbar\omega_c}{2}\frac{\partial}{\partial\phi}\nn\\
&&+q\frac{\tilde{\theta}Br_a^3}{c\mu_0\epsilon_1}\left[\frac{1}{r^2}+2\frac{\epsilon_1-\epsilon_2}{\epsilon_1+\epsilon_2+\epsilon_0}\frac{r}{r_b^3}\right]\cos\vartheta,
\label{Hamiltonian}
\eea

where $\omega_c=qB_0/m^*$  is the cyclotron frequency.

\subsection{Solving the Schr\"odinger Equation}\label{Sec_Solving_Scho_Ec}
In order to solve the time-independent Schrödinger equation  associated to the Hamiltonian shown in Eq. (\ref{Hamiltonian}), we adopt  spherical coordinates $\{r,\vartheta,\varphi\}$ for the wave function. Given that the Hamiltonian commutes with the component $\hat{L}_z=-i\hbar\,\partial_\varphi$ of the angular momentum, the familiar magnetic quantum number $m\in\mathbb{Z}$  is appropriate to label wave functions and energies. Furthermore,  any wave function admits a factorization in the following form,
\bea
\Psi_m(r,\vartheta,\varphi)=\frac{1}{r}\,\psi_m(r,u)\,e^{i\,m\varphi},
\label{Psi1}
\eea
with $u=\cos\vartheta$. The factor $r^{-1}$ is introduced for convenience. Moreover, the confinement of the QD inside the core-shell demands the boundary condition
\bea
\label{confinement}
\psi_m(r_a,u)=\psi_m(r_b,u)=0 .
\eea
It can be easily seen that $\psi_m$ satisfies the spectral problem
\bea
\label{eqpsi}
\hat{h}\,\psi_{m} = \mathcal{E}_{m}\,\psi_{m}
\eea
with eigenvalues 
\bea
 \mathcal{E}_{m} = E_m   + \frac{m\,\hbar\omega_c}{2} .
\eea
In Eq. (\ref{eqpsi}) the operator $\hat{h}$ is given by
\bea
\label{ReducedHamiltonian}
\hat{h}&=&-\frac{\hbar^2}{2m^{*}}\left[\frac{\partial^2}{\partial r^2}-\frac{1}{r^2}L_m^2\right]+V(r,u),
\eea
where
\bea
L_{m}^2=-\frac{\partial}{\partial u}\left[(1-u^2)\frac{\partial}{\partial u}\right]+\frac{m^2}{1-u^2}
\eea
and
\bea
\label{Potential}
V(r,u)=&&\dfrac{m^*}{2}\left[\omega^2+\frac{\omega_c^{2}}{4}(1-u^2)\right]r^2 \nn\\
&&+q\frac{\tilde{\theta}Br_a^3}{c\mu_0\epsilon_1}\left[\frac{1}{r^2}+2\frac{\epsilon_1-\epsilon_2}{\epsilon_1+\epsilon_2+\epsilon_0}\frac{r}{r_b^3}\right]u.\nn\\
\eea

To our best knowledge, the Schrödinger-like equation shown in Eq.~(\ref{eqpsi}) is not solvable by analytical means. To find its eigenvalues and eigenfunctions we employed one of the most accurate and efficient numerical methods, the Lagrange-mesh Method: an approximate  variational  method simplified  by  a  Gauss  quadrature  associated  with  the  mesh.  A detailed and complete description can be found in Ref.~\cite{BayeI}, while here we present  some of its main features as well as relevant formulas for a core-shell configuration.
  
 We take the function $\psi_m$ in the form
\bea
\psi_m(r,u)=\sum_{i=1}^{N_r}\sum_{j=0}^{N_u}c_{ij}^mf_i(r)g_j^m(u)
\label{Psim}
\eea
where $c_{ij}^m$ are real coefficients, with $N_r$ and $N_u$ sufficiently large. The function $f_i(r)$ is given by
\bea
f_i(r)&=&\frac{(-1)^{i+N_r+1}\sqrt{r_b-r_a}}{2}\frac{(r-r_a)(r_b-r)}{\sqrt{(r_i-r_a)(r_b-r_i)}}\times\nn\\ 
&&\frac{P_{N_r}\left(t(r)\right)}{r-r_i},
\eea
where $P_{N_r}(t(r))$ denotes the $N_r$-th Legendre polynomial, $t(r)$ is given by the formula
\bea
t(r)=\frac{2\,r}{r_b-r_a}+\frac{r_a+r_b}{r_a-r_b}
\eea
and $r_i$ are roots of $P_{N_r}(t(r))$, 
\bea
P_{N_r}(t(r_i))=0 ,\quad i=1,2,...,N_r\ .
\eea
Since $f_i(r_a)=f_i(r_b)=0$ for all $i$, the confinement condition shown in Eq. (\ref{confinement}) is fulfilled.

Following Ref. \cite{BayeIII}, it is convenient to take $g_j^m(u)$ in Eq.~(\ref{Psim}) as
\bea
g_{j}^m(u)= (-1)^{j+1}\left[\frac{(1-u_j^2)N_u!}{2(N_u+2|m|)!}\right]^{1/2}\frac{P_{N_u+|m|}^{|m|}(u)}{u-u_j},\nonumber\\
\eea
where $P_{N_u+|m|}^{|m|}(u)$ is the associated Legendre function with
\bea
P_{N_u+|m|}^{|m|}(u_j)=0\ ,\quad j=1,2,...,N_u\ .
\label{ujdef}
\eea

Needless to say, the roots $u_j$'s depend on the particular choice of $N_u$ and $m$. For the sake of simplicity we avoid to present such dependence explicitly. Coefficients $c_{ij}^m$ are determined by solving the  Lagrange-mesh system of equations in the Gauss quadrature approximation:
\bea
\label{secular}
\sum_{i=1}^{N_r}\sum_{j=1}^{N_u}\left\{ T_{ij,kl}^{m} + \left[V(r_i,u_j)  - \mathcal{E}_m\right]\delta_{ik}\delta_{jl}\right\}c_{ij}^{m} = 0\ .
\eea
 The kinetic matrix elements are  
\bea
T_{ij,kl}^{m} = \frac{1}{2}\left[t_{ik}\delta_{jl}\ +\ \frac{1}{r_i^2}s^{m}_{jl}\delta_{ik}\right]
\eea 
for which
\bea
\label{rij}
  t_{ik}=\frac{(-1)^{i+k} \left[(r_a +r_b)\left(r_i+r_k\right)-2 (r_i r_k+r_a r_b)\right]}{\left(r_i-r_k\right)^2 \sqrt{\left(r_i-r_a\right) \left(r_b-r_i\right) \left(r_k-r_a\right) \left(r_b-r_k\right)}}\nn\\
\eea
considering $i\neq k$ and
\bea
\label{ii}
t_{ii}=\frac{N_r(N_r+1)  \left(r_i-r_a\right) \left(r_b-r_i\right)+(r_a-r_b)^2}{3 \left(r_i-r_a\right)^2 \left(r_b-r_i\right)^2}
\eea
for $i=k$. In turn, for $j\neq l$
 \bea
   s^{m}_{jl} = (-1)^{j-l}\,\frac{2(1-u_j^2)^{1/2}(1-u_l^2)^{1/2}}{(u_j-u_l)^2},
\eea
while for $j= l$
\bea   s^{m}_{jj} = \frac{1}{3}(N_u+|m|)(N_u+|m|+1) + \frac{2(m^2-1)}{3(1-u_j^2)}.
\eea 

Further details can be found in Ref. \cite{BayeIII}. A particular case of formulas (\ref{rij}) and (\ref{ii}) was presented in Ref. \cite{BayeII} with $r_a=0$ and $r_b=1$. Here we give the extension of those formulas for a core-shell configuration.

Once the Eqs. (\ref{secular}) are solved for a given $m$, the low lying energies $\mathcal{E}_m$ as well as their corresponding coefficients $c_{ij}^{m}$ are determined and, ultimately, the approximate wave functions $\Psi_m$ is known. Numerical solutions of Eq. (\ref{secular}) were found with a computational code in \textit{Mathematica 12}. Numerical tests resulted in the conclusion that a mesh with $N_r=30$ and $N_u=30$ yields an excellent compromise with the desirable accuracy of at least 7 significant digits in energy and the expectation values.

\section{Optical Intersubband Properties}\label{Sec_Opt_Prop}

The optical absorption coefficient as well as the refractive index changes of the core-shell QD-TI are calculated by using the iterative density matrix formalism~\cite{Opticalproperties1,Opticalproperties2, Opticalproperties3}. We assume that the system is excited by an external electromagnetic plane wave of frequency $\omega_0$ and polarized in some $\boldsymbol{\hat{\eta}}$ direction,
\bea
{\bf E}_{\text{light}}(t)={\bf E}_0\cos(\omega_0 t)=\Tilde{\bf{E}}e^{i\omega_0 t}+\Tilde{\bf{E}}^*e^{-i\omega_0 t},
\eea
so that the linear and third-order perturbative nonlinear absorption coefficients are
\bea
\mathcal{A}^{(1)}(\omega_0)&=&\omega_0\sqrt{\frac{\mu}{\epsilon_r}}\left [ \frac{\sigma_v \hbar \Gamma_{ji}\left | M_{ji} \right |^2}{\left ( E_{ji}-\hbar \omega_0 \right )^2+ \left( \hbar  \Gamma_{ji} \right )^2} \right],
\eea
\bea
\mathcal{A}^{(3)}(\omega_0,I)&=&-\omega_0\sqrt{\frac{\mu}{\epsilon_r}}\left ( \frac{I}{2\epsilon_0 n_r c} \right )\nn\\
&\times& \frac{\sigma_v\hbar\Gamma_{ji}\left | M_{ji} \right |^2}{\left [ \left ( E_{ji}-\hbar \omega_0 \right )^2+ \left( \hbar  \Gamma_{ji} \right )^2 \right ]^2}\nn \\
&\times& \Bigg{\{}\frac{\left | M_{jj}-M_{ii} \right |^2\left [4E_{ji}\hbar \omega_0 -\hbar^2\left ( \omega_0^2-\Gamma_{ji}^2 \right ) \right ]}{E_{ji}^2+(\hbar\Gamma_{ji})^2}\nn\\
&-&\frac{3E_{ji}^2\left | M_{jj}-M_{ii} \right |^2 }{E_{ji}^2+(\hbar\Gamma_{ji})^2}+ 4\left | M_{ji} \right |^2\Bigg{\}},\nn\\
\eea
and the linear and the third-order nonlinear refractive index changes can be expressed as
\bea
\frac{\Delta n^{(1)}(\omega_0)}{n_r}&=&\frac{\sigma_v\left|M_{ji}\right|^2}{2n_r^2\epsilon_0}\left[\frac{E_{ji}-\hbar\omega_0}{(E_{ji}-\hbar\omega_0)^2+\left(\hbar\Gamma_{ji}\right)^2}\right],\nn\\
\frac{\Delta n^{(3)}(\omega_0)}{n_r}&=&-\frac{\sigma_v\left|M_{ji}\right|^2}{4n_r^3\epsilon_0}\frac{\mu c I}{\left[(E_{ji}-\hbar\omega_0)^2+\left(\hbar\Gamma_{ji}\right)^2\right]^2}\nn\\
&\times&\Bigg{\{}4(E_{ji}-\hbar\omega_0)\left|M_{ji}\right|^2-\frac{\left(M_{jj}-M_{ii}\right)^2}{\left(E_{ji}\right)^2+\left(\hbar\Gamma_{ji}\right)^2}\nn\\
&\times&\Bigg{[}\left(E_{ji}-\hbar\omega_0\right)\left[E_{ji}\left(E_{ji}-\hbar\omega_0\right)-\left(\hbar\Gamma_{ji}\right)^2\right]\nn\\
&-&\left(\hbar\Gamma_{ji}\right)^2\left(2E_{ji}-\hbar\omega_0\right)\Bigg{]}\Bigg{\}},
\eea
where $\mu$ is the permeability of the system defined as $\mu=1/\epsilon_0 c^2$, $\sigma_v$ is the carrier density and $I$ is the incident optical intensity defined as $I=2\sqrt{\epsilon_r/\mu}\vert \tilde{E}(\omega_0) \vert ^2$. Moreover, $\epsilon_r$ is the real part of the permittivity which is defined through $\epsilon_r=n^2_r\epsilon_0$ with $n_r$ the refractive index of the medium. The term $M_{ji}$ is the matrix element of the electric dipole moment defined as
\bea
M_{ji}=q\langle j|\left(\bf{r}\cdot\boldsymbol{\hat{\eta}}\right)|i\rangle
\label{dipolarmatrix},
\eea
where $\bf{r}$ is the position of the electron inside of the QD. On the other hand, the factor 
\bea
E_{ji}=E_j-E_i
\eea
is the energy difference between the $i$-th and $j$-th electronic levels, and $\hbar \omega_0$ the incident photon energy. Finally, the term $\Gamma_{ji}=1/\tau_{ji}$ is the relaxation rate defined through the relaxation time $\tau_{ji}$.

For the above, the total absorption coefficient and refractive index changes are given in this order of approximation by
\bea
\mathcal{A}(\omega_0,I) =\mathcal{A}^{(1)}(\omega_0)+\mathcal{A}^{(3)}(\omega_0,I),
\label{alpha_tot_def}
\eea
and
\bea
\frac{\Delta n(\omega_0)}{n_r}=\frac{\Delta n^{(1)}(\omega_0)}{n_r}+\frac{\Delta n^{(3)}(\omega_0)}{n_r}.
\eea
\subsection{Selection Rules}\label{Sec_Sel_Rul}
The dipolar matrix element of Eq.~(\ref{dipolarmatrix}) gives the allowed transitions of the system. Due to the spatial symmetry breaking induced by the external magnetic field, we consider two possible light polarizations: parallel and perpendicular to $\n{B}$, i.e., $\hat{\mathbf{e}}_z$ and $\hat{\mathbf{e}}_x$, respectively.

From Eq.~(\ref{Psi1}) the selection rule for the quantum number $m$ is given by
\bea
\label{selectionrule}
\Delta m=\left\{\begin{matrix}
0,\;\;\text{for}\;\;\hat{\mathbf{e}}_z\\
 \\
\pm 1,\;\;\text{for}\;\;\hat{\mathbf{e}}_x
\end{matrix}\right.,
\eea
and therefore, if the incident electric field is described by a polarization given by
\bea
\boldsymbol{\hat{\eta}}=\sin\beta\cos\gamma\,\hat{\mathbf{e}}_x+\sin\beta\sin\gamma\,\hat{\mathbf{e}}_y+\cos\beta\,\hat{\mathbf{e}}_z,
\eea
and $\beta,\gamma$ are the angles formed with the $z$- and $x$-axis, respectively, the non-vanishing dipole matrix elements are

\bea
\langle m_1,n_1|\mathbf{r}\cdot\boldsymbol{\hat{\eta}}|m_0,n_0\rangle&=&\sin\beta\cos\gamma\langle m_0\pm1,n_1|\hat{x}|m_0,n_0\rangle\nn\\
&+&\sin\beta\sin\gamma\langle m_0\pm1,n_1|\hat{y}|m_0,n_0\rangle\nn\\
&+&\cos\beta\langle m_0,n_1|\hat{z}|m_0,n_0\rangle,
\eea
which implies that any dipolar transition has no simultaneous information on the polarization parallel and perpendicular to $z$.

Two interesting situations occur when the second term in the potential $V(r,u)$ (proportional to $\tilde{\theta}$ and $B$) is absent, see Eq. (\ref{Potential}). Such situations appear in two cases:  (i) at zero magnetic field  and (ii) $\tilde{\theta}$=0.  For case (i) the Hamiltonian becomes spherically symmetric, therefore  hydrogen-type quantum numbers $(n,l,m)$ are good quantum numbers to label any state. For (ii), in addition to the quantum number $m$, parity $\nu=\pm1$ under the reflection $z\rightarrow-z$ is another good quantum number to label states. As a result, an additional selection rule to $ \Delta m=\pm1$ occurs for dipole transitions when $\boldsymbol{\hat{\eta}}=\hat{\mathbf{e}}_x$: only states with opposite parity have non-vanishing matrix dipole element.

In any case, the selection rules for all quantum numbers are implemented numerically in the framework of the Lagrange-mesh Method, see Appendix \ref{DipoleAppendix}. The Table~\ref{Tab:Sel_rule_x} shows the dipolar matix elements for the transition $\left|m,n\right.\rangle=\left|0,0\right.\rangle\rightarrow\left|1,n\right.\rangle$, normalized to the electron charge in units of the Bohr radius $a_0$. Such matrix elements are constructed for several values of the external magnetic field and $\tilde{\theta}$ normalized to the QED fine-structure constant $\alpha$, when the incident electromagnetic wave is polarized in the $x$-axis direction. As can be noted, the TI opens new transitions which otherwise are forbidden. 

\begin{table}[h]
\centering
\begin{tabular}{ |c|c|c|c|c| } 
\hline
$\tilde{\theta}/\alpha$& $\left|\langle m_0,n_0|\hat{x}|m_1,n_1\rangle\right|/a_0$ & $B=0$T &$B=3$T &$B=5$T \\
\hline
\multirow{3}{1em}{0} & $\left|\langle 0,0|\hat{x}|1,0\rangle\right|/a_0$ & 57.7125 & 57.6914 & 57.6538\\
& $\left|\langle 0,0|\hat{x}|1,1\rangle\right|/a_0$ & 0 & 0 & 0\\ 
& $\left|\langle 0,0|\hat{x}|1,2\rangle\right|/a_0$ & 0 & 0.0060 & 0.0367\\ 
& $\left|\langle 0,0|\hat{x}|1,3\rangle\right|/a_0$ & 0 & 0 & 0\\ 
& $\left|\langle 0,0|\hat{x}|1,4\rangle\right|/a_0$ & 0 & $<10^{-6}$ & $<10^{-6}$\\ 
\hline
\multirow{3}{1em}{11} & $\left|\langle 0,0|\hat{x}|1,0\rangle\right|/a_0$ & 57.7125 & 51.7673 & 48.1352\\
& $\left|\langle 0,0|\hat{x}|1,1\rangle\right|/a_0$ & 0 & 55.0625 & 53.8415\\ 
& $\left|\langle 0,0|\hat{x}|1,2\rangle\right|/a_0$ & 0 & 
0.6524& 0.9676\\ 
& $\left|\langle 0,0|\hat{x}|1,3\rangle\right|/a_0$ & 0 & 0.0250 & 0.0609\\ 
& $\left|\langle 0,0|\hat{x}|1,4\rangle\right|/a_0$ & 0 & 0.0006 & 0.0024\\ 
\hline
\multirow{3}{1em}{15} & $\left|\langle 0,0|\hat{x}|1,0\rangle\right|/a_0$ & 57.7125 & 49.6414 & 45.8171\\
& $\left|\langle 0,0|\hat{x}|1,1\rangle\right|/a_0$ & 0 & 55.7048 & 47.9628\\ 
& $\left|\langle 0,0|\hat{x}|1,2\rangle\right|/a_0$ & 0 & 0.8514 & 1.0352\\ 
& $\left|\langle 0,0|\hat{x}|1,3\rangle\right|/a_0$ & 0 & 0.0432 &0.0857\\
& $\left|\langle 0,0|\hat{x}|1,4\rangle\right|/a_0$ & 0 & 0.0013 & 0.0045\\ 
\hline
\end{tabular}
\caption{Dipolar matrix elements in units of Bohr radius $a_0$ and normalized to the electron charge for several values of the magnetic field $B$ and $\tilde{\theta}$ normalized to the QED fine-structure constant $\alpha$.}
    \label{Tab:Sel_rule_x}
\end{table}

\begin{figure}[h]
    \centering
    \includegraphics[scale=0.4]{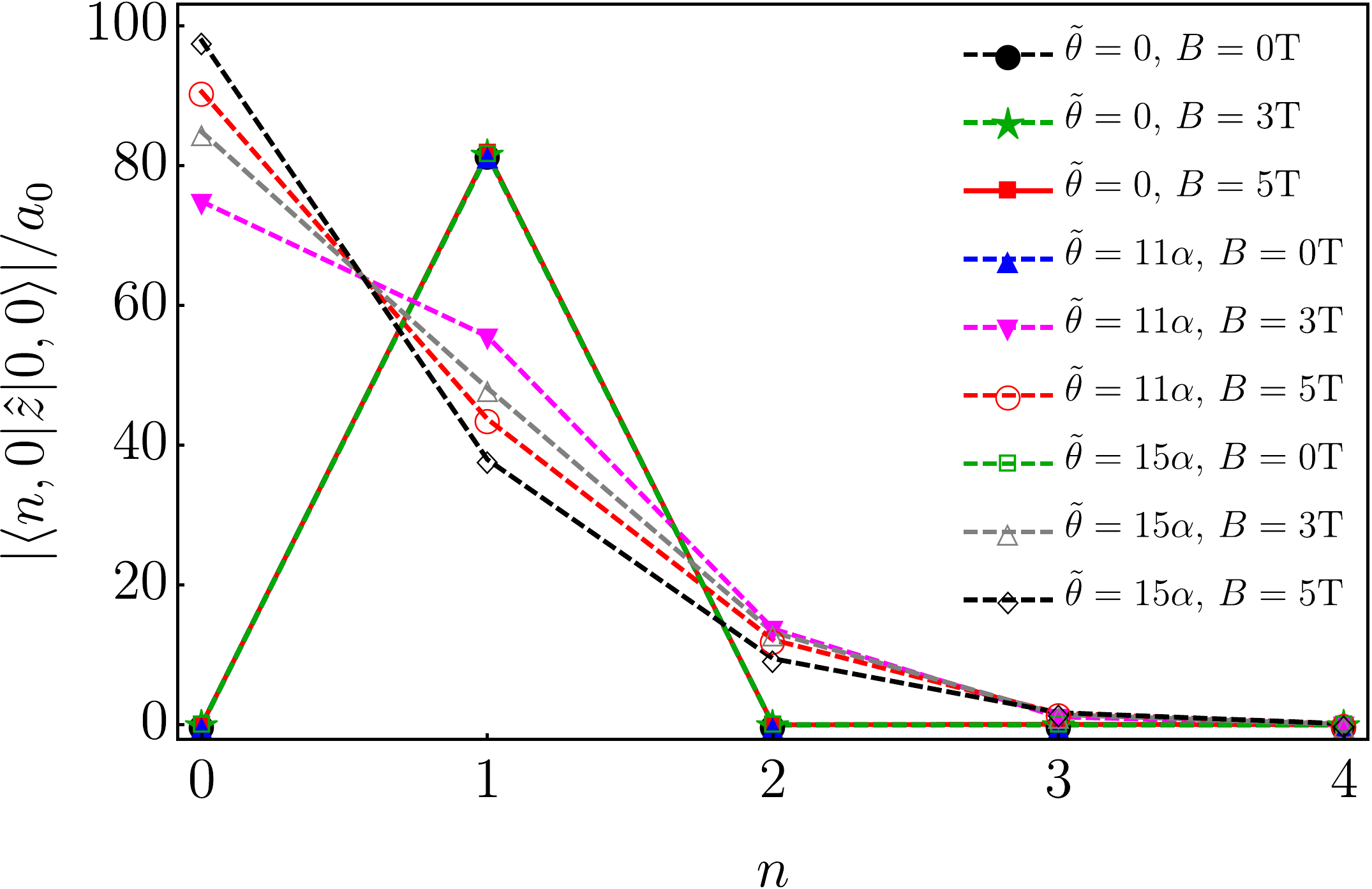}
    \caption{Dipolar matrix elements in units of Bohr radius $a_0$ and normalized to the electron charge for several values of the magnetic field $B$ and $\tilde{\theta}$.}
    \label{Fig:Sel_rul_z}
\end{figure}

The selection rules for the transition $\left|m,n\right.\rangle=\left|0,0\right.\rangle\rightarrow\left|0,n\right.\rangle$ when $\boldsymbol{\hat{\eta}}\parallel\mathbf{B}$ are shown in Fig.~\ref{Fig:Sel_rul_z} for several values of $B$ and $\Tilde{\theta}$. 

In the same way as perpendicular polarization, the presence of the TI promotes the appearance of new electronic transitions. Moreover, the transition probabilities when $\boldsymbol{\hat{\eta}}\parallel\mathbf{B}$ have a maximum at the lowest quantum number implying that the system prefers to remain in the same state. Thus, we encourage a complete study of all allowed transitions which can be good experimental signals to infer the TME nature of nanoscopic arrangements.

\section{Results and Discussion}\label{Sec_results}
In order to compute the optical properties of our system, we set the parameters as follows: $m^{*}$=0.067$m_0$ as the effective electron mass for a GaAs QD where $m_0$ is the free electron mass; $r_a$=5 nm, $r_b$=10 nm, and $r_c=$200 nm as the radii of the TI, the external GaAs surface, and the current loop, respectively. Furthermore, we have chosen $V_0=$36 meV as the value of the confining potential, $\epsilon_2/\epsilon_0\equiv\epsilon_r=n_r^2=12.53$ as the relative permittivity of GaAs, and we fix the relaxation rate to $\Gamma_{ji}=(0.2\,\text{ps})^{-1}$ as a typical value for a GaAs QD. The carrier density is taken as $\sigma_v=3\times 10^{22}$ m$^{-3}$. Concerning the material constants for the TI, we choose TlBiSe$_{2}$ as reference, therefore, $\epsilon_1=4\epsilon_0$, $\mu_1=\mu_0$ which yields a value of $\tilde{\theta}\in[\alpha,11\alpha]$ \cite{TlBiSe_2}.  However, in order to sharpen the effects, we will increase $\tilde{\theta}$ until 15$\alpha$.

\begin{figure}[h]
    \centering
    \includegraphics[scale=0.4]{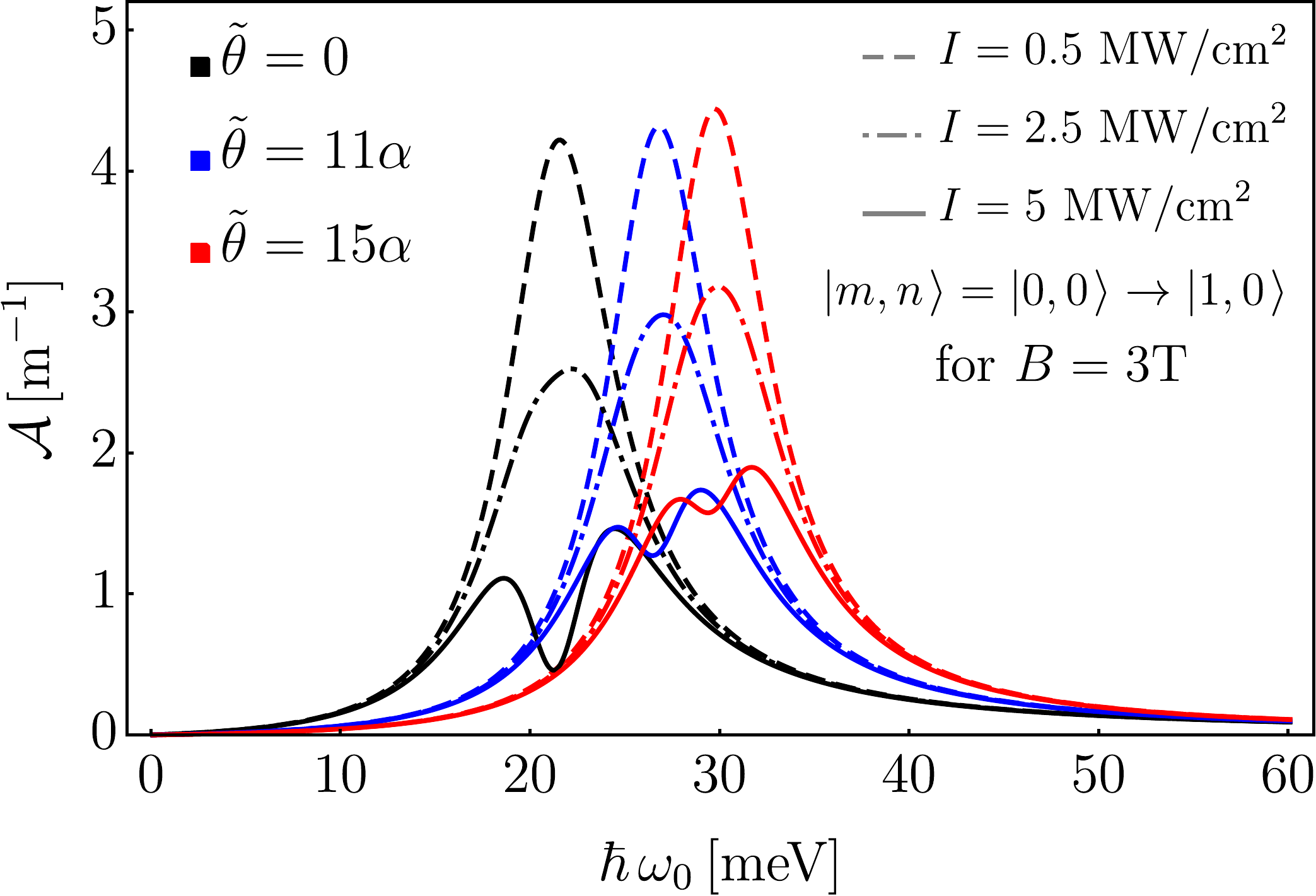}
    \caption{Total absorption coefficient defined in Eq.~(\ref{alpha_tot_def}) for the transition $\left|m,n\right.\rangle=\left|0,0\right.\rangle\rightarrow\left|1,0\right.\rangle$ with several values of the intensity $I$ and $\tilde{\theta}$ with $B=3$T and $\boldsymbol{\hat{\eta}}=\hat{\mathbf{e}}_x$.}
    \label{Fig:Abs_several_theta_B3T}
\end{figure}

Figure~\ref{Fig:Abs_several_theta_B3T} shows the total absorption coefficient defined in Eq.~(\ref{alpha_tot_def}) as a function of the incident photon energy for the transition $\left|m,n\right.\rangle=\left|0,0\right.\rangle\rightarrow\left|1,0\right.\rangle$ for different values of $\tilde{\theta}$ and the incident optical intensity $\mathcal{I}$, and are considered at constant magnetic field $B=3$T. The figure is constructed by assuming a light polarization oriented along the $x$-axis direction. As can be noted, the optical absorption shifts towards higher energies with increasing $\tilde{\theta}$. Although the functional behaviour of $\mathcal{A}$ remains unaltered, the effect of the topological insulator is to increase the absorption and move the energetic optical response region by $\hbar\omega_0\sim15$meV. Also, at lower intensities the light absorption increases, whereas for high values the non-linear effects become more important, and therefore, given the relative sign between $\mathcal{A}^{(1)}$ and $\mathcal{A}^{(3)}$, the peak intensity decreases.
\begin{figure}[h]
    \centering
    \includegraphics[scale=0.4]{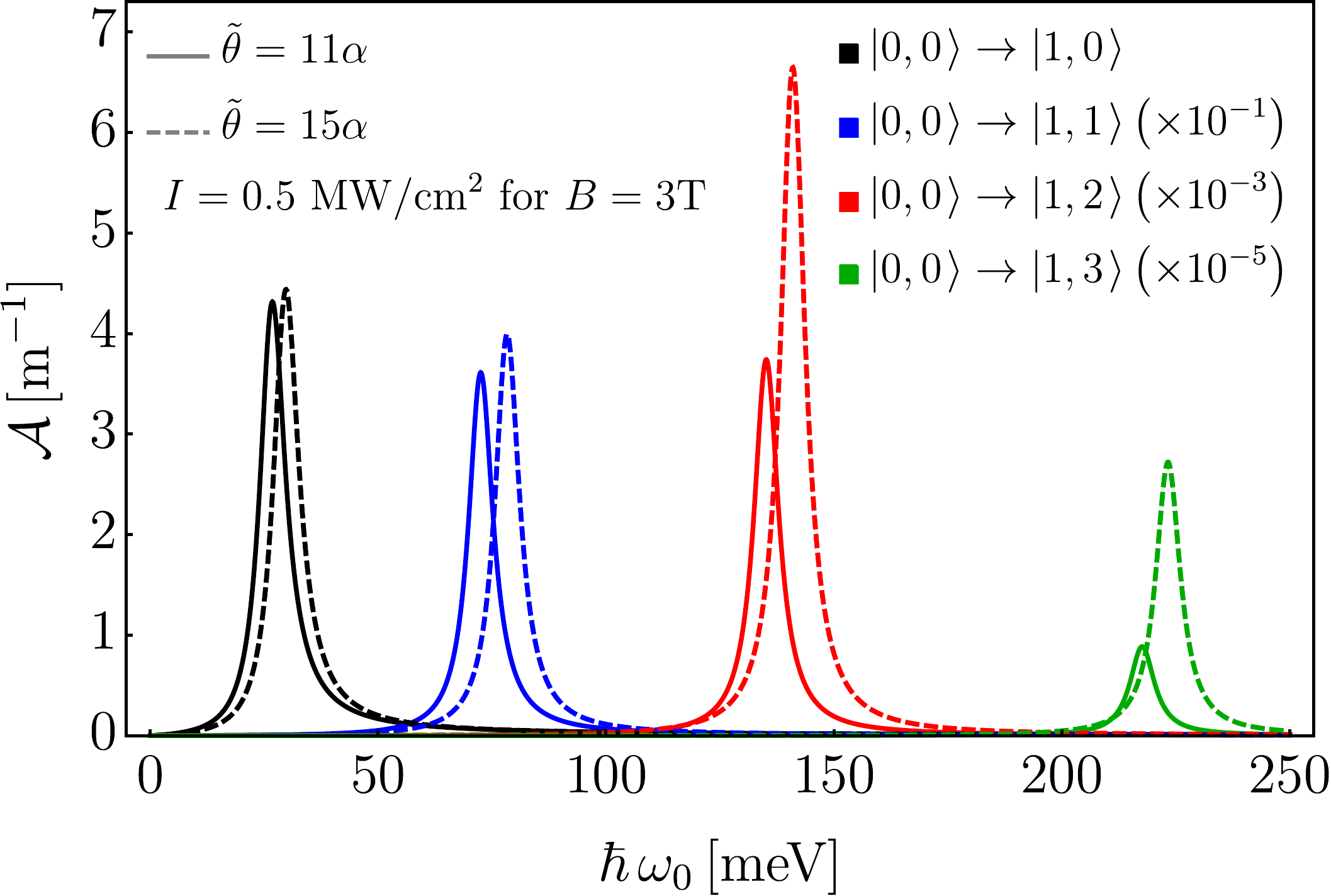}
    \caption{Total absorption coefficient as function of the photon energy for several allowed electronic transitions when $\tilde{\theta}=11\alpha, 15\alpha$, $\boldsymbol{\hat{\eta}}=\hat{\mathbf{e}}_x$ and fixed values of $I$ and $B$.}
    \label{Fig:Abs_two_theta}
\end{figure}

In order to assess the impact of the TI in the full absorption spectrum, Fig.~\ref{Fig:Abs_two_theta} shows the optical absorption coefficient for the lower energetic transitions, i.e., $\left|m,n\right.\rangle=\left|0,0\right.\rangle\rightarrow\left|1,i\right.\rangle$ with $0\leq i\leq 4$. The magnetic field has been set to $B=3$T, the incident intensity to $I=0.5$MW/cm$^2$ and $\boldsymbol{\hat{\eta}}=\hat{\mathbf{e}}_x$. Recalling the results of Table~\ref{Tab:Sel_rule_x}, it is clear that transitions to higher quantum numbers have less probability to take place. Thus, the system prefers to absorb a photon that modifies its angular quantum number for 1 and leaves the other quantum numbers unaltered. It is worth mentioning that the increasing value of $\tilde{\theta}$ intensifies the optical absorption, which is more evident when the system goes to higher principal quantum numbers.
\begin{figure}[h]
    \centering
    \includegraphics[scale=0.39]{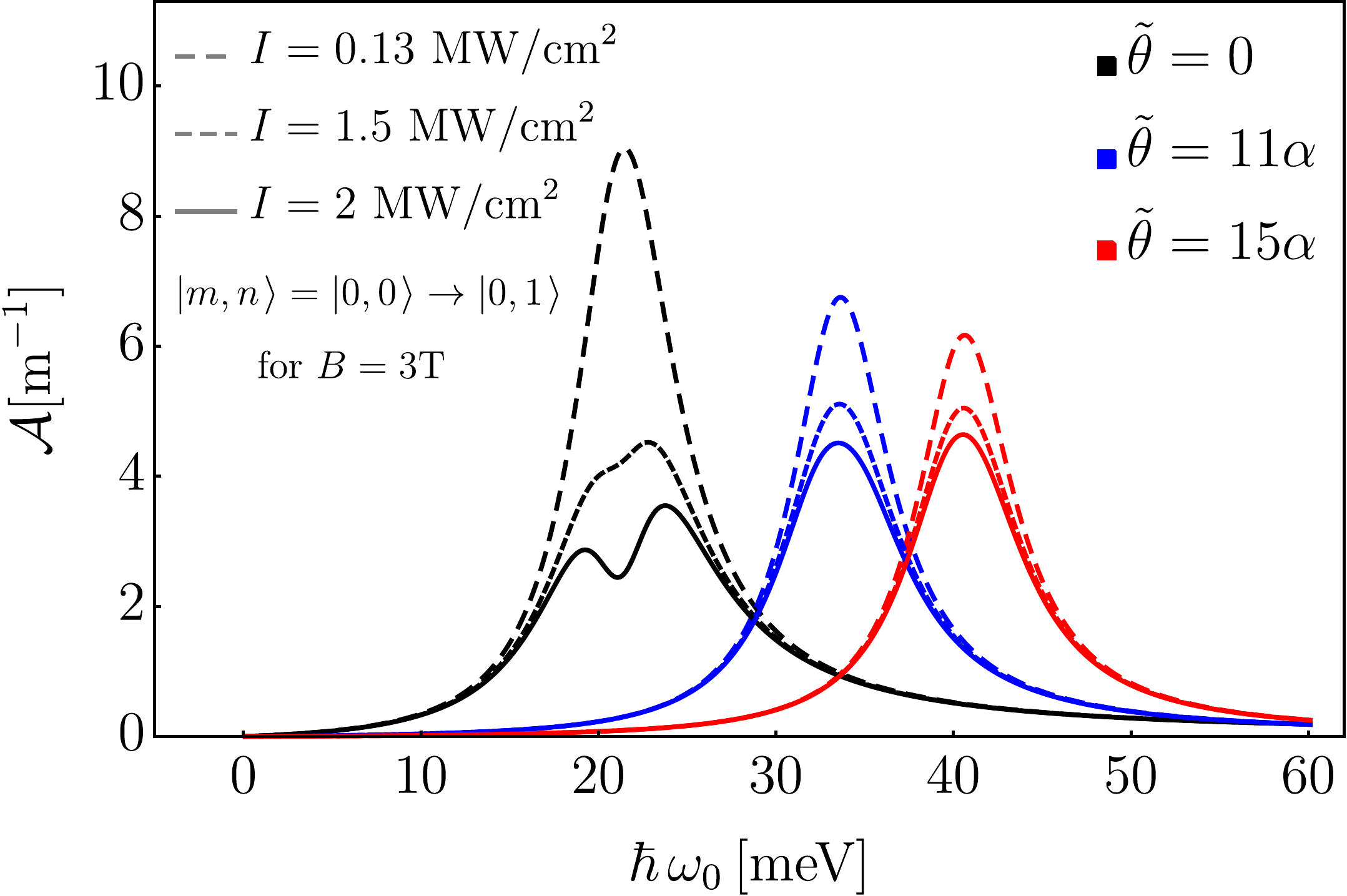}
    \caption{Total absorption coefficient defined in Eq.~(\ref{alpha_tot_def}) for the transition $\left|m,n\right.\rangle=\left|0,0\right.\rangle\rightarrow\left|0,1\right.\rangle$ with several values of the intensity $I$ and $\tilde{\theta}$ when $B=3$T and $\boldsymbol{\hat{\eta}}=\hat{\mathbf{e}}_z$.}
    \label{Fig:Abs_several_z}
\end{figure}

For the sake of completeness, Fig.~\ref{Fig:Abs_several_z} shows the same information presented in Fig.~\ref{Fig:Abs_several_theta_B3T} but with $\boldsymbol{\hat{\eta}}=\hat{\mathbf{e}}_z$. Contrary to the latter an increasing value of $\tilde{\theta}$ implies a lower light absorption. Also, the non-linear effects become more evident at smaller values of the optical intensity. This behavior can be understood in terms of the electron's dynamics when the monochromatic light source is applied to the QD. For a non-vanishing magnetic field, three forces are acting on the free charge: the dipolar term $q\mathbf{E}_{\text{light}}$, the Lorentz force $q\mathbf{v}\times\mathbf{B}$, and the magnetoelectric field of Eq.~(\ref{ElectricField}) which is $\theta$-dependent. If the light polarization is oriented along the $x$-axis, the competition with the other forces makes a resonant oscillation difficult. Now, if the polarization is $z$-oriented, the only force which rivals the one due to the external driving source is the force of the magnetoelectric field, whose orientation enhances the oscillation along the z-axis.


\begin{figure}[h]
    \centering
    \includegraphics[scale=0.39]{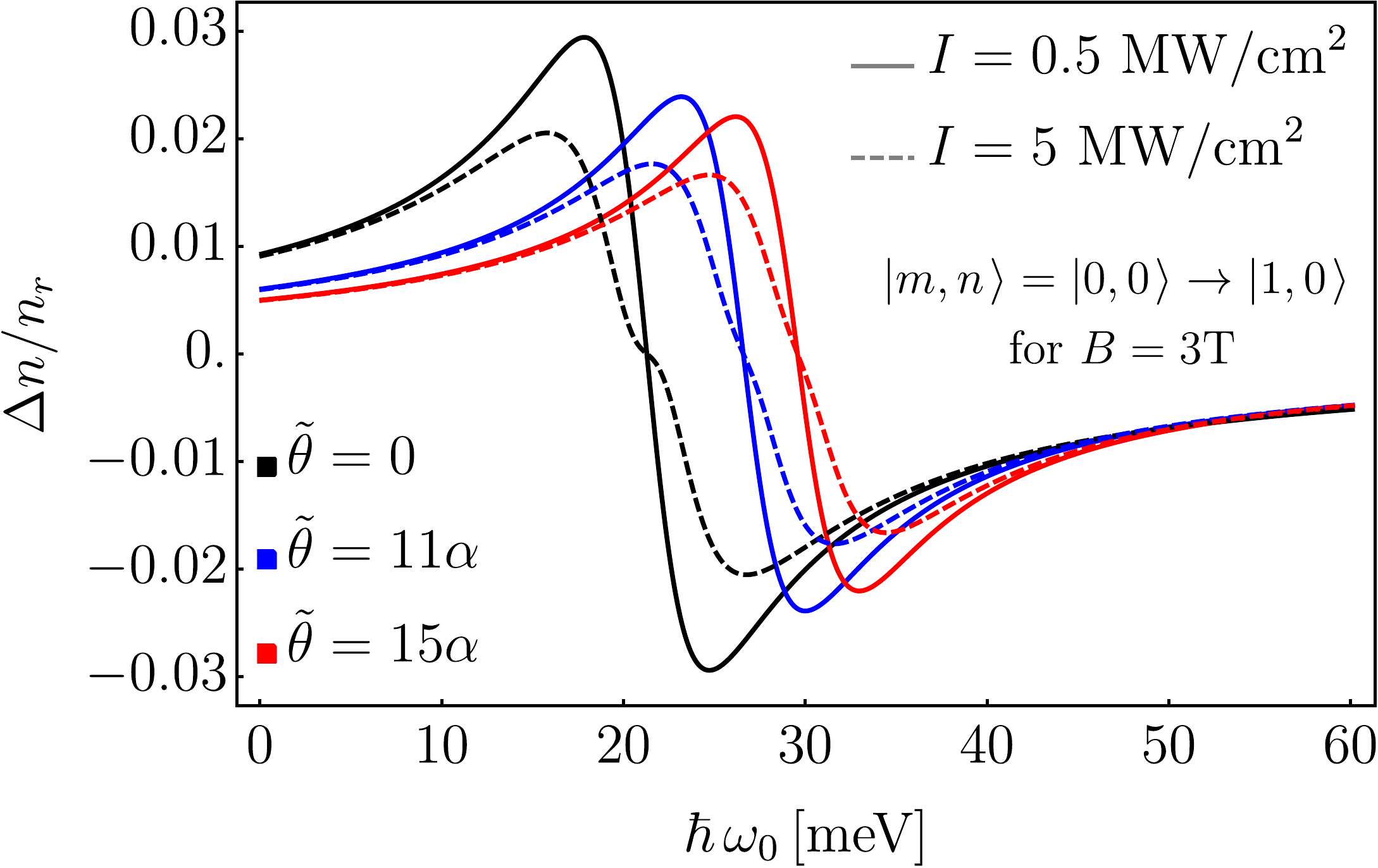}
    \caption{Refraction index changes as a function of the photon energy for several values of $\tilde{\theta}$ and $I$ for $B=3$T in the transition $\left|m,n\right.\rangle=\left|0,0\right.\rangle\rightarrow\left|1,0\right.\rangle$. The light polarization is taken as $\boldsymbol{\hat{\eta}}=\hat{\mathbf{e}}_x$.}
        \label{Fig:Ref_Ind_chang_1}
\end{figure}
\begin{figure}[h]
    \centering
    \includegraphics[scale=0.39]{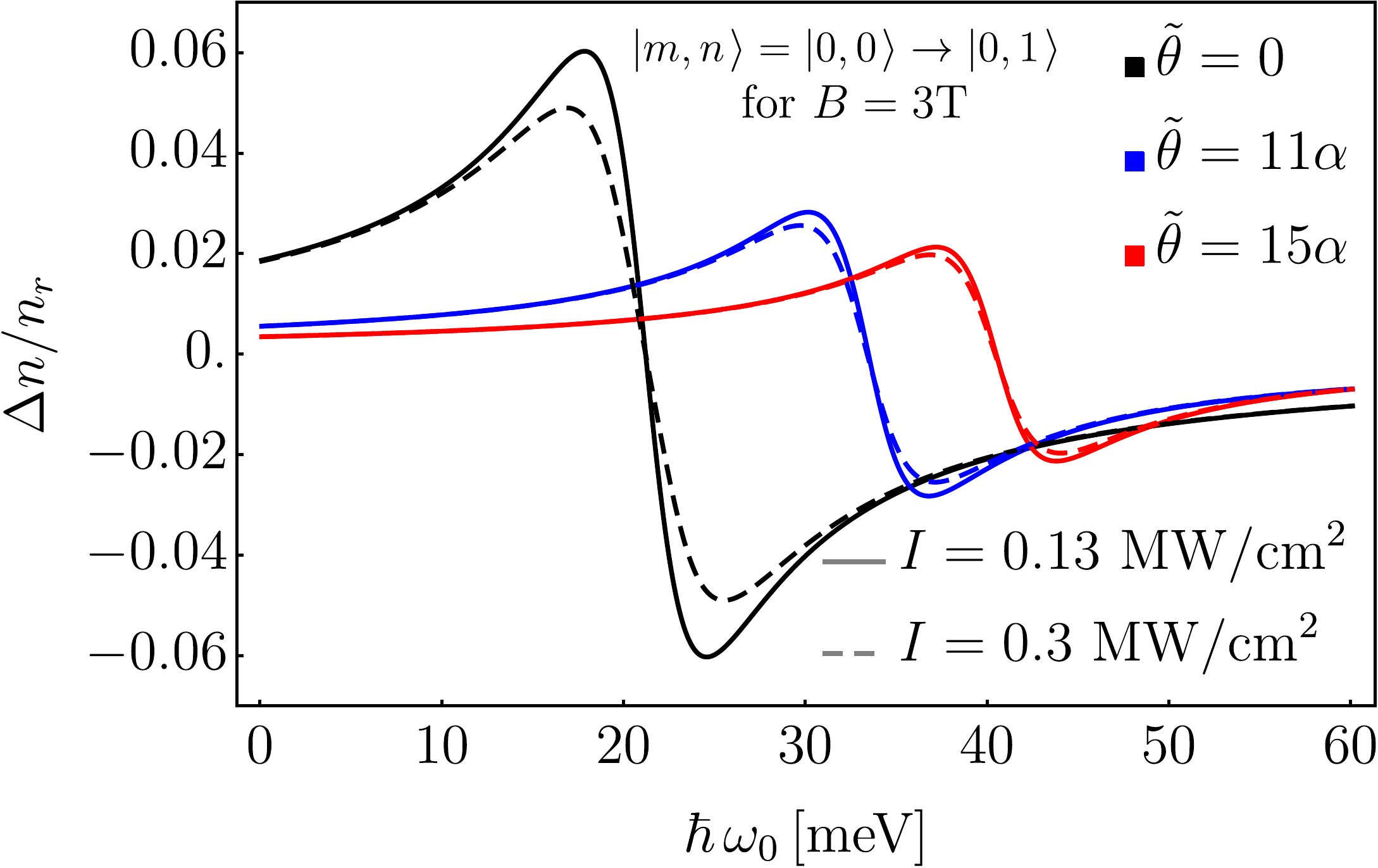}
    \caption{Refraction index changes as a function of the photon energy for several values of $\tilde{\theta}$ and $I$ for $B=3$T in the transition $\left|m,n\right.\rangle=\left|0,0\right.\rangle\rightarrow\left|0,1\right.\rangle$. The light polarization is taken as $\boldsymbol{\hat{\eta}}=\hat{\mathbf{e}}_z$.}
    \label{Fig:Ref_Ind_chang_pol_z}
\end{figure}

Finally, Figs.~\ref{Fig:Ref_Ind_chang_1} and~\ref{Fig:Ref_Ind_chang_pol_z} shows the total refraction index changes when the incident light has polarizations $\boldsymbol{\hat{\eta}}=\hat{\mathbf{e}}_x$ and $\boldsymbol{\hat{\eta}}=\hat{\mathbf{e}}_z$, respectively. These plots are constructed as a function of the photon energy for different values of $\tilde{\theta}$ and the optical intensity $I$ for $B=3$T and take into account the allowed transition from the ground state to the first excited energy level. The influence of $\tilde{\theta}$ is to reduce the changes in $n(\omega)$ regardless of the electromagnetic wave polarization. The radiation intensity has an appreciable impact when $\boldsymbol{\hat{\eta}}=\hat{\mathbf{e}}_x$ and the non-linear effects become more evident as the intensity parameter grows. Conversely, if the light intensity has a polarization parallel to the magnetic field, then it doesn't significantly affect the refractive index changes.

\section{Summary and Conclusions}\label{Sec_Concl}
In this work, we have studied the optical intersubband properties of an in-shell GaAs quantum dot coupled to a topological insulator which is described by the so-called $\theta$-electrodynamics. In order to account for the magnetoelectric response of the TI and calculate the electric and magnetic fields induced in the GaAs, the Green's Function method has been used to find the 4-potential $A^\mu$. The Schrödinger equation related to the vector and scalar potentials is solved by the Lagrange-mesh method, which allows for calculating the eigenvalues and eigenvectors of the system, and therefore, the dipolar matrix elements dictating the selection rules. Finally, the optical intersubband properties, i.e., the total absorption coefficient and the total refractive index changes are computed in the iterative density matrix formalism to non-linear third order.

The results show that the presence of the topological insulator allows for a new set of dipolar transitions, which depend on the light polarization parallel or parallel or perpendicular to the external magnetic field. As a general result, we find that an arbitrary light polarization $\boldsymbol{\hat{\eta}}$ promotes transitions with $\Delta m=0$ (associated with $\boldsymbol{\hat{\eta}}=\hat{\mathbf{e}}_z$) or $\Delta m=\pm1$ (associated with $\boldsymbol{\hat{\eta}}=\hat{\mathbf{e}}_x$), which implies that certain dipolar transitions are generated by the projection of $\boldsymbol{\hat{\eta}}$ either along the z-axis or in the xy-plane, but not simultaneously. If $\boldsymbol{\hat{\eta}}=\hat{\mathbf{e}}_x$, such transitions have their maxima when the magnetic quantum number $m$ changes one unit, whereas the system prefers to not perform a dipolar transition if $\boldsymbol{\hat{\eta}}=\hat{\mathbf{e}}_z$. Those transition amplitudes play a crucial role in the optical properties and they are sensitive to changes in $\tilde{\theta}$.

The absorption coefficient, as well as the refractive index changes, are modified substantially due to the TI's presence. For instance, depending on the direction of the incident light polarization, the total absorption function enhances or decreases, compared with the absence of such magnetoelectric material, i.e., when $\tilde{\theta}=0$. On the other hand, the induced magnetic and electric fields reduce the changes in the refractive index. By computing those optical properties for dipolar transitions beyond the ground state to the first excited state, we find that a higher value of $\tilde{\theta}$ intensifies the photon absorption. Clearly, given the magnetoelectric nature of the TI, all these results depend on the value of the external magnetic field (or current intensity in the external loop) for a fixed magnetoelectric coupling $\tilde{\theta}$. However, large effects are expected to occur if the magnetic field is adequately increased. 

Comments on the fact that real systems may not respect the full spherical symmetry and its impact on the optical properties \cite{Li-Chen,McBride} should be pointed out. Arguably, such symmetry breaking can be regarded as a perturbation, which leads to two scenarios. The first one, when the perturbation comes from a geometric potential $V_p$ modeling the spherical defects of the system, such that our work can be regarded as the non-perturbed system. Only if the realistic geometry modeled by the potential $V_R$ is close to the spherical one $W_0$, i.e. $|V_R-W_0|\ll|W_0|$, and after performing standard perturbation theory with unperturbed states $|n^{(0)}\rangle$ in spherical geometry, then the energy spectrum at first order will be $\Delta E\sim\langle n^{(0)}|V_p|n^{(0)}\rangle$. Within this point of view and regarding the selection rules of Eq.~(\ref{selectionrule}), the dipolar matrix elements should not be far from the values reported in this manuscript. Thus, the absorption peaks size will be slightly modified. The second scenario is posed when the problem is tackled from the beginning in its proper geometry, for example, oblate or prolate spheroidal coordinates. In this perspective, there are well-known studies in literature providing evidence that perturbation theory cannot predict correctly the new spectrum when the original geometry is changed. For instance, the hydrogen atom confined in two kind of cavities: $(i)$ a spherical cavity of radius $R$~\cite{Baye-Sen}, and $(ii)$ an ellipsoidal cavity characterized by the semi-axes $a=b\neq c$~\cite{Kang-Liu et al}. This scenario leads us to recalculate unavoidably all the relevant functions and quantities presented in this manuscript, which will constitute future work.

Finally, we want to remark that the $\theta$-effects are proportional to the QED fine-structure constant ($\alpha=1/137$), which implies a challenging difficulty in the direct measure of phenomena arising from the TME such as induced torques \cite{Maghrebi}, magnetic monopole effective terms \cite{Qi Science}, atomic energy shifts \cite{Alejo} or even the ordinary magnetoelectric effect in Cr$_2$O$_3$ \cite{ME measurements}, to mention some examples. Nevertheless, in the spirit of an indirect measurement as proposed in Refs. \cite{Okada,Wu,Dziom}, the existence of dipolar transitions and the sensitivity of the transition amplitudes to changes in theta, reflected in the optical properties, may be of experimental interest, because they offer an alternative path to infer the $\theta$-parameter (MEP) by optical measurements. Other coupled configurations are being currently explored and will be reported soon.

\section*{ACKNOWLEDGMENTS}
The authors acknowledge financial support from Consejo Nacional de Ciencia y Tecnolog\'ia (CONACyT). O. J. Franca acknowledges partial support for this work from CONACyT Project No. 237503 and Project \# IN103319 from Direcci\'on General de Asuntos del Personal Acad\'emico (Universidad Nacional Aut\'onoma de M\'exico). J. C. del Valle was supported by a CONACyT Ph. D. Grant No.570617 (México). The authors also thank Dr. Christine Gruber for a thorough reading of the manuscript and many helpful comments.

\section*{AUTHOR CONTRIBUTIONS}
J. D. Castaño-Yepes proposed the system and model, performed the optical coefficients calculations, defined the relevant observables, discussed the results, wrote the Secs.~\ref{Sec_hamiltonian},~\ref{Sec_Opt_Prop},~\ref{Sec_results},~\ref{Sec_Concl} and coordinated the project.

C. F. Ramírez-Gutierrez proposed the system and model, ran the calculations of eigenvalues and dipolar elements, discussed the results.

O. J. Franca performed the $\theta$-ED calculations, wrote the Sec.~\ref{Sec:ThetaED} and their appendixes.

J. C. del Valle developed the Lagrange-mesh method computational code, wrote the Sec.~\ref{Sec_Solving_Scho_Ec} and its Appendix.

All the authors contributed to Sec.~\ref{sec:intro} and reviewed the manuscript.

\appendix
\section{Calculation of the 4-potential $A^{\mu}$}\label{CalculoAmu}
In this appendix we will determine the 4-potential $A^\mu(\mathbf{x})$ of a constant magnetic field in the presence of a spherical TI of radius $r_a$ which is embedded in a concentric dielectric sphere of radius $r_b$ (See Fig. \ref{Esquema}) using the GF method. Performing the convolution between the GF (\ref{GFM}) and the current density (\ref{j}) through Eq. (\ref{A and GF}), we have
\bea
A^{\mu}(\mathbf{x})&=&\mathcal{I}r_c\int^{2\pi}_{0}d\phi' \sin\phi'G^{\mu}_{\;\;1}\left(r,\vartheta,\phi;r_c,\frac{\pi}{2},\phi'\right)\nonumber\\
&&-\mathcal{I}r_c\int^{2\pi}_{0}d\phi'\cos\phi'G^{\mu}_{\;\;2}\left(r,\vartheta,\phi;r_c,\frac{\pi}{2},\phi'\right)\;,\nn\\
\eea
where we have already computed the integrals involving Dirac delta functions. To carry out the azimuthal integral, we rewrite $\sin\phi'$ and $\cos\phi'$ in terms of complex exponentials in the GF (\ref{GFM}). Then, we use the explicit form of spherical harmonics \cite{Jackson}, and find
\bea
A^{\mu}(\mathbf{x})&=&4\pi^{2}\mathcal{I}r_c\sum^{\infty}_{l=0}\sum^{l}_{m=-l}Y_{l}^{m}(\vartheta,\phi)Y_{l}^{1}\left(\frac{\pi}{2},0\right)\nn\\
&&\times\left[ig^{\mu}_{lm1;\;\;1}(r,r_c)+g^{\mu}_{lm1;\;\;2}(r,r_c)\right]\nn\\
&&+4\pi^{2}\mathcal{I}r_c\sum^{\infty}_{l=0}\sum^{l}_{m=-l}Y_{l}^{m}(\vartheta,\phi)Y_{l}^{-1}\left(\frac{\pi}{2},0\right)\nn\\
&&\times\left[-ig^{\mu}_{lm-1;\;\;1}(r,r_c)+g^{\mu}_{lm-1;\;\;2}(r,r_c)\right],\label{A mu}\nn\\
\eea
where we have already used the orthonormality of the Fourier basis \cite{Arfken}.\\

Although we observe that $A^3(\mathbf{x})\equiv0$ from Eq. (\ref{A mu}) and the reduced GF equations (\ref{red GFM}), we have to calculate each component separately because there is no compact expression for the reduced GF $g^\mu_{lmm';\nu}$. This will be shown in the following subsections.\\

\subsection{$A^{0}(\mathbf{x})$ component}
Setting $\mu=0$ in Eq. (\ref{A mu}), we have
\bea
\Phi(\mathbf{x})&=&4\pi^{2}\mathcal{I}r_c\sum^{\infty}_{l=0}\sum^{l}_{m=-l}Y_{l}^{m}(\vartheta,\phi)Y_{l}^{1}\left(\frac{\pi}{2},0\right)\nn\\
&&\times\left[ig^{0}_{lm1;\;\;1}(r,r_c)+g^{0}_{lm1;\;\;2}(r,r_c)\right]\nn\\
&&+4\pi^{2}\mathcal{I}r_c\sum^{\infty}_{l=0}\sum^{l}_{m=-l}Y_{l}^{m}(\vartheta,\phi)Y_{l}^{-1}\left(\frac{\pi}{2},0\right)\nn\\
&&\times\left[-ig^{0}_{lm-1;\;\;1}(r,r_c)+g^{0}_{lm-1;\;\;2}(r,r_c)\right].\nn\label{A0 almost 1}\\
\eea

Now, we need to analyze explicitly the square bracket terms. From Eq. (\ref{red GFM}) we read
\begin{eqnarray}
ig^{0}_{lm1;\;\;1}(r,r_c)&+&g^{0}_{lm1;\;\;2}(r,r_c)\nn\\
&=&r_a\tilde{\theta}\sqrt{l(l+1)}S_{l}^{(\epsilon,1)}(r,r_c)\delta_{m0},\nn\\
-ig^{0}_{lm-1;\;\;1}(r,r_c)&+&g^{0}_{lm-1;\;\;2}(r,r_c)\nn\\
&=&-r_a\tilde{\theta}\sqrt{l(l+1)}S_{l}(r,r_c)\delta_{m0},
\label{sum red GFs}
\end{eqnarray}
where we used the well-known properties of ladder operators $\hat{L}_{\pm}=\hat{L}^x\pm i\hat{L}^y$ \cite{Sakurai}. Substituting Eq. (\ref{sum red GFs}) in Eq. (\ref{A0 almost 1}), we find
\begin{eqnarray}
\Phi(\mathbf{x})&=&4\pi^{2}\tilde{\theta}\mathcal{I}r_ar_c\sum^{\infty}_{l=0}\sqrt{l(l+1)}S_{l}^{(\epsilon,1)}(r,r_c)Y_{l}^{0}(\vartheta,\phi)\nn\\
&&\times\left[Y_{l}^{1}\left(\frac{\pi}{2},0\right)-Y_{l}^{-1}\left(\frac{\pi}{2},0\right)\right]\;.\label{A0 almost 2}
\end{eqnarray}

Then, using the next identity for spherical harmonics \cite{Jackson},
\bea
&&Y_{l}^{1}\left(\frac{\pi}{2},0\right)-Y_{l}^{-1}\left(\frac{\pi}{2},0\right)\\\label{dif 1}
&=&\left\{
\begin{array}{lll}
0 & ,\;l=2n,\;n\in\mathbb{N}\\
\sqrt{\frac{4n+3}{\pi (2n+1)(2n+2)}}f(n) & ,\;l=2n+1,\;n\in\mathbb{N}\nn
\end{array}\right.,
\eea
where
\begin{equation}\label{f(n)}
f(n)=(-1)^{n+1}\frac{\Gamma(n+3/2)}{\Gamma(n+1)\Gamma(3/2)}\;,
\end{equation}
we rewrite Eq. (\ref{A0 almost 2}) as shown 
\bea
\Phi(\mathbf{x})&=&\frac{4\pi^{2}\tilde{\theta}\mathcal{I}r_ar_c}{\sqrt{\pi}}\sum^{\infty}_{n=0}\sqrt{4n+3}f(n)\nn\\
&&\times S_{2n+1}^{(\epsilon,1)}(r,r_c)Y_{2n+1}^{0}(\vartheta,\phi).
\eea

In the limit $r\ll r_c$, which guarantees a constant and uniform magnetic field in the direction of $\mathbf{\hat{e}}_z$, it is enough to take only the leading term $n=0$ \cite{Jackson}. Thus, 
\begin{eqnarray}
\Phi(\mathbf{x})&=&\frac{4\pi^{2}\tilde{\theta}\mathcal{I}r_ar_c}{\sqrt{\pi}}\sqrt{3}f(0)S_{1}^{(\epsilon,1)}(r,r_c)Y_{1}^{0}(\vartheta,\phi)\;,\nonumber\\
&=&-6\pi\tilde{\theta}\mathcal{I}r_ar_c\cos\vartheta\nn\\
&&\times\frac{\mathcal{F}_1^{(\epsilon)}(r,r_a)\mathcal{F}^{(1)}_1(r_a,r')}{1+2r_a^2\T{\theta}^{2}\mathcal{F}^{(1)}_1(r_a,r_a)\mathcal{F}^{(\epsilon)}_1(r_a,r_a)},\nn\\
\end{eqnarray}
where we have already substituted Eq. (\ref{S r r'}) in the last equation.\\

Noting that for TIs, $\tilde{\theta}\sim\alpha\gg\alpha^2\sim\tilde{\theta}^{2}$, and recalling that one of the largest values measured for $\tilde{\theta}$ is 0.22 for TbPO$_4$ \cite{TbPO4}, an ordinary magnetoelectric, we can use this value to estimate $\tilde{\theta}^{2}\leq4.8\times10^{-2}=\tilde{\theta}^{2}_{\text{TbPO}_{4}}$. Thus, we can neglect the quadratic terms in $\tilde{\theta}^{2}$. Considering this, we find the final expression for $\Phi(\mathbf{x})$ at first order in $\tilde{\theta}$, which is
\bea
\Phi(\mathbf{x})=\frac{6\pi\T{\theta}r_a^3\mathcal{I}}{\epsilon_1r_b^3r_c}\left(r_c^3-r_b^3\right)\left[\frac{1}{r^2}+\frac{2\left(\epsilon_1-\epsilon_2\right)}{\epsilon_1+\epsilon_2+1}\frac{r}{r_b^3}\right]\cos\vartheta\;,\nn\\
\eea
which is expressed in Gaussian units. After using the conversion  factors presented in \cite{Jackson} and restoring the vacuum speed of light $c=1/\sqrt{\epsilon_0\mu_0}$ and $\hbar$ to their values in SI units we obtain
\begin{equation}
\Phi(\n{x})=\frac{\tilde{\theta}Br_a^3}{c\mu_0\epsilon_1}\left[\frac{1}{r^2}+2\frac{\epsilon_1-\epsilon_2}{\epsilon_1+\epsilon_2+\epsilon_0}\frac{r}{r_b^3}\right]\cos\vartheta\;,
\end{equation}
which is just Eq. (\ref{Phi SI}).

\subsection{$A^{i}(\mathbf{x})$ components}
Setting $\mu=1$ in Eq. (\ref{A mu}), we obtain
\bea
A^{1}(\mathbf{x})&=&4\pi^{2}\mathcal{I}r_c\sum^{\infty}_{l=0}\sum^{l}_{m=-l}Y_{l}^{m}(\vartheta,\phi)Y_{l}^{1}\left(\frac{\pi}{2},0\right)\nn\\
&&\times\left[ig^{1}_{lm1;\;\;1}(r,r_c)+g^{1}_{lm1;\;\;2}(r,r_c)\right]\nn\\
&&+4\pi^{2}\mathcal{I}r_c\sum^{\infty}_{l=0}\sum^{l}_{m=-l}Y_{l}^{m}(\vartheta,\phi)Y_{l}^{-1}\left(\frac{\pi}{2},0\right)\nn\\
&&\times\left[-ig^{1}_{lm-1;\;\;1}(r,r_c)+g^{1}_{lm-1;\;\;2}(r,r_c)\right].\nn\label{A1 almost 1}\\
\eea

Analyzing explicitly the square bracket terms from Eq. (\ref{red GFM}) we have
\begin{eqnarray}
ig^{1}_{lm1;\;\;1}(r,r_c)+g^{1}_{lm1;\;\;2}(r,r_c)
&=&i\mathcal{F}_{l}^{(1)}(r,r_c)\delta_{m1}\nonumber,\label{sum red GFs A1}\\
-ig^{1}_{lm-1;\;\;1}(r,r_c)+g^{1}_{lm-1;\;\;2}(r,r_c)
&=&-i\mathcal{F}_{l}^{(1)}(r,r_c)\delta_{m1}\;,\nonumber\\
\end{eqnarray}
where we have used the well-known properties ladder operators \cite{Sakurai} and neglected the $\tilde{\theta}^{2}$ terms as discussed in the previous section of this appendix.\\

Substituting the Eqs. (\ref{sum red GFs A1}) in Eq. (\ref{A1 almost 1}), we arrive at 
\begin{eqnarray}
A^{1}(\mathbf{x})&=&i4\pi^{2}\mathcal{I}r_c\sum^{\infty}_{l=0}\mathcal{F}^{(1)}_l(r,r_c)\left[Y_{l}^{1}(\vartheta,\phi)Y_{l}^{1}\left(\frac{\pi}{2},0\right)\right.\nn\\
&&\left.-Y_{l}^{-1}(\vartheta,\phi)Y_{l}^{-1}\left(\frac{\pi}{2},0\right)\right]\;.\label{A1 almost 2}
\end{eqnarray}

Then, using the following expression obtained by definition of spherical harmonics \cite{Jackson},
\begin{eqnarray}
&&Y_{l}^{1}(\vartheta,\phi)Y_{l}^{1}\left(\frac{\pi}{2},0\right)-Y_{l}^{-1}(\vartheta,\phi)Y_{l}^{-1}\left(\frac{\pi}{2},0\right)
\nonumber\\
&&=\left\{
\begin{array}{lll}
0 & ,l=2n\\
i\frac{4n+3}{2\pi (4n^{2}+6n+2)}f(n)P^{1}_{2n+1}(\cos\vartheta)\sin\phi & ,l=2n+1\\
\end{array}\right.,\nonumber\\
\end{eqnarray}
where $n\in\mathbb{N}$ and $f(n)$ was defined in Eq. (\ref{f(n)}). We rewrite Eq. (\ref{A1 almost 2}) as
\begin{eqnarray}
A^{1}(\mathbf{x})&=&-2\pi \mathcal{I}r_c\sin\phi\sum^{\infty}_{n=0}\frac{4n+3}{4n^{2}+6n+2}f(n)\nn\\
&&\times\mathcal{F}^{(1)}_{2n+1}(r,r_c)P^{1}_{2n+1}(\cos\vartheta)\;.
\end{eqnarray}

Once again in the limit $r\ll r_c$, which guarantees a constant and uniform magnetic field in the direction of $\mathbf{\hat{e}}_z$, the leading term is $n=0$ \cite{Jackson}, so we find the final expression for $A^{1}(\mathbf{x})$ as
\bea
A^{1}(\mathbf{x})&=&3\pi \mathcal{I}r_c\mathcal{F}^{(1)}_l(r,r_c)\sin\vartheta\sin\phi\;,\nn\\
&=&\frac{3\pi \mathcal{I}}{r_b^3 r_c}
\left(r_b^3-r_c^3\right)r\sin\vartheta\sin\phi\;,\label{A1}
\eea
where we have already substituted Eq. (\ref{free GF 2}) and 
$r_b^{3}-r_c^{3}<0$ because $r_b\ll r_c$.\\

Analogously, the expression for $A^{2}(\mathbf{x})$ is
\begin{equation}\label{A2}
A^{2}(\mathbf{x})=\frac{3\pi \mathcal{I}}{r_b^3 r_c}
\left(r_c^{3}-r_b^{3}\right)r\sin\vartheta\sin\phi,
\end{equation}
where $r_c^{3}-r_b^{3}>0$ because $r_b\gg r_c$.\\

From Eqs. (\ref{A1}) and (\ref{A2}) the vector potential found here is
\begin{equation}
\mathbf{A}(\mathbf{x})=\frac{3\pi \mathcal{I}}{r_b^{3}r_c}(r_c^{3}-r_b^{3})r\sin\vartheta\,\mathbf{\hat{e}}_\phi\;,\label{A vec 1}
\end{equation}
which is expressed again in Gaussian units. After using the conversion  factors presented in \cite{Jackson} and restoring the vacuum speed of light $c=1/\sqrt{\epsilon_0\mu_0}$ and $\hbar$ to their values in SI units we obtain
\bea
\mathbf{A}(\mathbf{x})=\frac{B r}{2}\sin\vartheta\,\mathbf{\hat{e}}_\phi,
\eea
where we identified
\begin{equation}\label{B0 appendix}
B=\frac{3\mu_0\mathcal{I}}{2r_b^{3}r_c}(r_c^{3}-r_b^{3}),
\end{equation}
which is just Eq.~(\ref{vec_pot}) , i.e., the vector potential for a constant and uniform magnetic field in the direction $\mathbf{\hat{e}}_z$ of standard ED with a magnitude $B$ given by Eq.~(\ref{B0 appendix}).


\section{Dipole Elements}\label{DipoleAppendix}

In the framework of the Lagrange-mesh method we (approximately) estimated  the matrix elements of the dipole operator given in Eq.~(\ref{dipolarmatrix}). To calculate them, it is enough to compute the matrix elements of the position operator
\bea
\langle m_1,n_1|\mathbf{r}|m_0,n_0\rangle.
\eea

For simplicity we assume same values for $N_r$ and $N_u$ for the states $| m_1,n_1\rangle$ and $|m_0,n_0\rangle$, respectively. In the Gauss quadrature approximation the integration required by the matrix element for the $z$-direction ($\Delta m=0$)  results in the computation of a finite sum 
\bea
\label{zdipole}
|\langle m_0,n_1|z|m_0,n_0\rangle|&&=\nonumber\\ 
&&\left|\sum _{i=1}^{N_r}r_i \sum _{j=1}^{N_u}  u_j\,c_{ij}^{(m_0,n_1)}\,  c_{ij}^{(m_0,n_0)}\right|.\nn\\
\eea 
In Eq. (\ref{zdipole}) $c_{ij}^{(m_k,n_l)}$ denotes the coefficient $c_{ij}^{m_k}$ for the excitation $n_l$.

Following from cylindrical symmetry of the system, it is immediate to check that
\bea
|\langle m_1,n_1|x|m_0,n_0\rangle|=|\langle m_1,n_1|y|m_0,n_0\rangle|.
\eea

The integration required for this matrix element is done in the Gauss quadrature approximation and thus it is the result of a finite sum,
\bea
&&|\langle m_1,n_1|x|m_0,n_0\rangle|= \nonumber\\ 
&&\frac{1}{2}  \left|\sum _{i=1}^{N_r}r_i\sum _{j=1}^{N_u} \sum _{k=1}^{N_u} c_{ij}^{(m_0,n_0)}c_{ik}^{(m_1,n_1)}\,\mathcal{W}_j^{m_1}(u_k^{m_0})\right.\nn\\
&\times&\left.\sqrt{1-\left(u_k^{m_0}\right)^2}\sqrt{\lambda_k^{m_0}}\right|,
\eea
where the weights of the $u$-mesh are given by
\bea
\label{weight}
\lambda_k^{m_0}=\frac{2 (2|m_0|+N_u)!}{N_u ! \left[1-\left(u_k^{m_0}\right)^2\right]\mathcal{W}_k^{m_0}\left(u_k^{m_0}\right)},
\eea
and we define
\bea
\mathcal{W}_j^m(u)=\frac{d}{du}g_j^m(u),
\eea
recalling that $u_k^{m_0}$ is the root of Eq.~(\ref{ujdef}) when $m=m_0$.
Here the Gauss quadrature in the $u$-mesh defined by  $(N_u,m_0)$ has been used for integration; a similar expression occurs when the mesh is defined by $(N_u,m_1)$.

\end{document}